\documentclass[11pt]{article}

\usepackage{graphicx,amssymb,amsmath,amsfonts}
\usepackage{hyperref}
\usepackage[margin=1.2in]{geometry}
\usepackage[noadjust]{cite}

\numberwithin{equation}{section}

\usepackage[utf8]{inputenc}

\date{\today}

\def\be{\begin{equation}}
\def\ee{\end{equation}}

\newcommand{\avg}[1]{\langle #1 \rangle}

\newcommand{\alp}{\alpha^{\prime}}

\newmuskip\pFqmuskip

\newcommand*\pFq[6][8]{%
  \begingroup 
  \pFqmuskip=#1mu\relax
  \mathchardef\normalcomma=\mathcode`,
  \mathcode`\,=\string"8000
  \begingroup\lccode`\~=`\,
  \lowercase{\endgroup\let~}\pFqcomma
  {}_{#2}F_{#3}{\left(\genfrac..{0pt}{}{#4}{#5}\Big | #6\right)}%
  \endgroup
}
\newcommand{\pFqcomma}{{\normalcomma}\mskip\pFqmuskip}

\newcommand{\pmat}{\begin{pmatrix}}
\newcommand{\fpmat}{\end{pmatrix}}
\newcommand{\eq}{\begin{equation}}
\newcommand{\feq}{\end{equation}}
\newcommand{\cas}{\begin{cases}}
\newcommand{\fcas}{\end{cases}}

\newcommand{\eqarray}{\begin{eqnarray}}
\newcommand{\feqarray}{\end{eqnarray}}

\newcommand{\bq}{\bullet\quad}

\begin{document}

\begin{titlepage}
\title{\textbf{Measuring chaos in string scattering processes}\vspace{10mm}}

\author{\textbf{Massimo Bianchi}\(^{[a,b]}\) \\
\href{mailto:massimo.bianchi@roma2.infn.it}{massimo.bianchi@roma2.infn.it} \and
\textbf{Maurizio Firrotta}\(^{[a,b]}\) \\
\href{mailto:maurizio.firrotta@gmail.com }{maurizio.firrotta@gmail.com} \and
\textbf{Jacob Sonnenschein}\(^{[c]}\) \\ \href{mailto:cobi@tauex.tau.ac.il}{cobi@tauex.tau.ac.il} \and \textbf{Dorin Weissman}\(^{[d]}\) \\ \href{mailto:dorin.weissman@apctp.org}{dorin.weissman@apctp.org}
}

\date{\(^{[a]}\)\emph{Dipartimento di Fisica, Università di Roma Tor Vergata},\\
\emph{Via della Ricerca Scientifica 1, 00133, Roma, Italy} \\[1.3\baselineskip] 
\(^{[b]}\)\emph{INFN sezione di Roma Tor Vergata} \\
\emph{Via della Ricerca Scientifica 1, 00133 Roma, Italy} \\[1.3\baselineskip] 
\(^{[c]}\)\emph{The Raymond and Beverly Sackler School of Physics and Astronomy},\\
\emph{Tel Aviv University, Ramat Aviv 69978, Tel Aviv, Israel}
\\ [1.3\baselineskip]
 \(^{[d]}\)\emph{Asia Pacific Center for Theoretical Physics},\\
	\emph{Pohang University of Science and Technology, Pohang 37673, Republic of Korea} 
 \\[1.5\baselineskip] \today}
	
\maketitle

\begin{abstract}
We analyze the amplitudes of one highly excited string (HES) state with two or three tachyons in open bosonic string theory. We argue that these processes are chaotic by showing that the spacing ratios of successive peaks in the angular dependence of the amplitudes are distributed as predicted by the $\beta$-ensemble of random matrix theory (RMT). We show how the continuous parameter $\beta$ depends on the level and helicity of the scattered HES state. We derive the scattering amplitude of an HES and three tachyons and show that it takes the form of the Veneziano amplitude times a dressing factor, and that the dressing is chaotic as a function of the scattering angle, in the sense that its spacing ratios match with RMT predictions.
\end{abstract}

\end{titlepage}

\flushbottom

\tableofcontents

\clearpage

\section{Introduction} \label{sec:intro}
Chaotic processes are common in a wide range of domains from physics, chemistry, and biology, to sociology and more. In physics they appear both in classical and in quantum phenomena, both for single-body and for many-body systems. 

Several definitions and measures have been proposed in the analysis of chaos. In the context of quantum Hamiltonian systems one time-tested method is to compare the statistics of the energy spectrum to the predictions of Random Matrix Theory (RMT), namely, to matrices whose elements are randomly chosen from a (usually Gaussian) random distribution (see \cite{Mehta:book,Akemann:book} and many references therein). This correspondence was verified for single-particle systems  as well as for many-body systems. Most of the latter models were discrete ones and only recently studies of continuous systems, and in particular quantum field theory (QFT), have been performed \cite{Srdinsek:2020bpq}. The toolkit for analyzing QFT models is not yet complete and devising new and efficient measures is an important task in the field.

Recently in \cite{Bianchi:2022mhs}, motivated by \cite{Rosenhaus:2020tmv}, we proposed a novel measure for quantum scattering processes. The crucial idea was to relate the angular distribution of scattering amplitudes to the spectral statistics of quantum systems governed by RMT. We believe that this measure could be applied on any quantum amplitude given by an $S$-matrix.

In  \cite{Bianchi:2022mhs} we applied it in particular to scattering on a leaky torus \cite{Gutzwiller:1983} and to the decay of highly excited string (HES) states into two tachyons. A comparison was made against the ``prototype'' of quantum chaos that is the distribution of non-trivial zeros of the Riemann zeta function \cite{Berry:1986,Odlyzko:1987}.

It is very natural to suspect that not only the decays of HES states, but the scattering processes of such states admit chaotic behavior as well. This study was initiated in \cite{Rosenhaus:2020tmv, Gross:2021gsj, Rosenhaus:2021xhm}, where scattering amplitudes involving three or four external legs were analyzed and were shown to display erratic behavior. In particular a drastic difference was observed in the  scattering amplitudes where the scattered HES state is changed in a minor way in its structure. However, the authors of \cite{Rosenhaus:2020tmv, Gross:2021gsj, Rosenhaus:2021xhm} did not directly quantify these type of differences. The measure that we proposed in \cite{Bianchi:2022mhs} was introduced to fulfill this purpose. Indeed it allowed us to demonstrate that decays of HES states into two tachyons did admit a chaotic behavior. 

The goal of this paper is to expand on the results of \cite{Bianchi:2022mhs} and to apply the method also to four-point scattering amplitudes involving an HES state. Chaotic scattering of HES states is  especially interesting  because  of the proposed correspondence with black holes \cite{Horowitz:1996nw, Damour:1999aw}. Determining chaotic patterns of string scattering together with the knowledge about the chaotic nature of black holes, may shed new light on the string/black hole correspondence.

Let us briefly review the relation of the spectral structure of a Hamiltonian system with eigenvalues $E_n$ and RMT. Consider the spacings between eigenvalues \cite{Berry:1977wk},
\begin{equation} \delta_n = E_{n+1} - E_{n} \end{equation}
and  define the \emph{ratios of successive spacings}
\begin{equation} \label{eq:rn} r_n \equiv \frac{E_{n+1}-E_n}{E_n-E_{n-1}} = \frac{\delta_{n+1}}{\delta_n} \end{equation}
In chaotic systems the level spacings are distributed as the spacings of eigenvalues of random matrices. In order to see this one must first ``unfold'' the spectrum \cite{Bohigas:1983er}, to account for the average density of states and expose the erratic fluctuations. The related distribution function for $r_n$, as a ratio of nearby spacings is not very sensitive to the unfolding procedure and as such can be used directly on the spectrum. Distributions of these spacing ratios has been successfully used as a measure of chaos in works such as \cite{Huse:2007,Srdinsek:2020bpq}. In some cases the normalized ratios \(\tilde r_n \equiv min\{r_n,\frac{1}{r_n}\}\), defined to be between 0 and 1, are used.

In analogy to the energy spacings, we proposed in \cite{Bianchi:2022mhs} to analyze  the spacings between successive peaks of a scattering amplitude ${\cal A(\alpha)}$ as a function of a continuous kinematical variable $\alpha$, relevant to the scattering process under scrutiny. As for the energy spacings, we define the spacings \(\delta_n\) between successive peaks and their ratios $r_n$, and compare the resulting distribution to RMT predictions. 

Our analysis of the chaotic behavior of decays and scattering processes involving HES states can be divided in three steps. First, we construct the HES state in the DDF approach (after Del Giudice, Di Vecchia and Fubini) \cite{DelGiudice:1971yjh}.\footnote{See also \cite{Hindmarsh:2010if, Skliros:2011si, Bianchi:2019ywd, Addazi:2020obs, Aldi:2019osr, Aldi:2020qfu, Aldi:2021zhh, Firrotta:2022cku} for more recent reviews and other applications.}
In the second step we compute the relevant decay and scattering amplitudes as a function of the available kinematical variables. In this paper, this is done for a decay of an HES state into two tachyons and for the four point amplitude of an HES state and three tachyons. In both cases the amplitude is analyzed as a function of an angle. In the former case 
the angle is the difference between the emission angle of the outgoing tachyons and the momentum of the photons used to create the DDF state. In the latter it is the usual scattering angle in the $2\to2$ process.
The third  step is to determine the locations of the maxima of the amplitudes, then compute the adjacent spacings and their ratios. We then perform a statistical analysis of the probability distribution function of the ratios for various levels $N$ and spin/helicity $J$ of the HES state. This is fitted to the predicted distribution of $r_n$ from the Gaussian $\beta$-ensemble \cite{Atas:2013dis}, in which $\beta$ is a continuous variable interpolating between the classical Gaussian ensembles - orthogonal (GOE), unitary (GUE) and symplectic (GSE) - of RMT.
  
The results of our analysis are the following:
\begin{itemize}
\item
For the decay processes we find that the distribution of spacings of peaks of the amplitude is well modelled by the RMT formula of the $\beta$-ensemble, with the parameter $\beta$ depending on the level $N$ and the helicity $J$ of the HES state. In the range of $N = 50$--$1600$, we find that $\beta$ is decreasing from $3.4$ to around $1.7$. This is also observed as a slow monotonous increase of the measured average $\langle r_n \rangle$ as a function of $N$.
\item
For the four point scattering amplitude, we show that the amplitude is given by the Veneziano amplitude times a chaotic dressing factor which depends on the HES state. We analyze these dressing factors in the high-energy fixed-angle limit and the Regge limit, for HES states with $N=100$ and find similar distributions for their $r_n$ with values of $\beta$ around 2 (the GUE value).
\item
In some cases, one can see clearly a transition from chaotic to regular spacings as one moves from small to large scattering angles.
\item
The chaotic behavior is observed for generic HES states, but it completely disappears for for states in the leading Regge trajectory or nearby states, i.e. for states with $N\approx J$.
\end{itemize} 

The paper is organized as follows. After this introduction, in section \ref{sec:measures} we present and discuss measures of chaos in quantum scattering amplitudes. In particular we describe the novel measure that we have proposed in \cite{Bianchi:2022mhs} and mention certain generalizations. In section \ref{sec:beta} we briefly review the $\beta$-ensemble of random matrices, including its Coulomb gas description. Section \ref{sec:HES} is devoted to a review of highly excited string states. We describe the construction of HES states using DDF operators and discuss integer partitions and the role of the helicity $J$. We then study  the chaotic behavior of the decay of these highly excited states in section \ref{sec:3pt}. We write down the decay amplitude of an HES to two tachyons and perform a statistical analysis of the relevant spacing ratios. This includes specifying a prescription of selection of the states and the fitting model. We then present the result of this analysis. Section \ref{sec:4pt} is devoted to chaotic four point scattering process involving one HES and three tachyons. The scattering amplitude of these processes is derived and written down. The amplitude  is then analyzed in the high-energy fixed-angle regime and the Regge limit. We determine the spacing ratios for the four-point scattering amplitude, and describe the chaotic behavior in the two high energy limits. We summarize the paper and mention several open questions in section \ref{sec:summary}.

For the benefit of the reader we add four appendices. In appendix \ref{app:4ptkin} we describe the kinematics of the four-point amplitude at hand. We then present a derivation of the HES-three tachyon amplitude in appendix \ref{app:4ptderivation}. In appendix \ref{app:lognormal} we make an explicit comparison of the $beta$-ensemble with the log-normal distribution of $r_n$ which we utilized as our main fitting model in our previous work \cite{Bianchi:2022mhs}. Random partitions of a large integer $N$ are discussed in appendix \ref{app:randompartitions}.
 
\section{Measure of chaos for quantum scattering amplitudes} \label{sec:measures}
Quantum scattering processes are characterized by a scattering amplitude. Suppose first that the scattering amplitude is a function of a scattering angle $\alpha$, i.e. of a single continuous real parameter.

In analogy to energy level spacings and their ratios, we propose to analyze a scattering amplitude ${\cal A}(\alpha)$ as a function of a scattering angle $\alpha$ as follows. In our case the ``eigenvalues'' would be the positions of local maxima and/or minima of the amplitude. We usually utilize the logarithmic derivative
\be F(\alpha) \equiv \frac{d}{d\alpha}\log {\cal A}.\ee
to find them. Then, our discrete levels are given by the set of zeros of \(F(\alpha)\) in the range $(0,2\pi)$:\footnote{Symmetries of the amplitude in the angle would allow us to eventually reduce the range to  $(0,\pi/2)$.}
\be F(z_n) = 0 \ee
Then we define, as for the energy levels, the spacings \(\delta_n\) and the ratios of consecutive spacings, \(r_n\) and $\tilde r_n$:
\be \delta_n = z_n - z_{n+1}\ee
and
\be r_n \equiv \frac{z_{n+1}-z_n}{z_n-z_{n-1}} = \frac{\delta_{n+1}}{\delta_n}\,,\qquad \tilde r_n =min\{r_n,\frac{1}{r_n}\} \ee

As mentioned in the introduction, when comparing level spacings to the predictions of random matrix theory, one typically has to perform an unfolding of the spectrum. A prototypical and useful example which we discussed in some detail in \cite{Bianchi:2022mhs} is the distribution of spacings of non-trivial zeros of the Riemann zeta function. In that case, the $z_n$ are the solutions of $F(z_n) = \zeta(\frac12 + i z_n) = 0$. There are infinitely many solutions, but one can consider finite subsets of them by limiting the range of $z_n$. Then, the normalized spacings $\overline\delta_n = (z_n-z_{n-1})\frac{\log(z_n/2\pi)}{2\pi}$ are known to match almost exactly with the distribution of spacings in the Gaussian unitary ensemble. In this case the formula for unfolding is known explicitly, but one could start with $r_n$ as defined above to see the agreement with the GUE without normalizing the spacings, since the logarithmic dependence in $\overline\delta_n$ is slow enough such that $r_n \approx \overline\delta_{n+1}/\overline\delta_n$. In many practical applications, unfolding is done by fitting the average density to a polynomial rather than a logarithm. Still, as long as this function is slowly varying at the scale of the spacings, the distribution of $r_n$ is only weakly affected by the unfolding procedure.

A generic scattering amplitude would depend on several kinematic variables like the incoming scattering angle, impact parameters, momenta, etc. In particular, a $2\to2$ scattering amplitude can be written as a function of two independent Mandesltam variables $s$ and $t$, or alternatively as a function of $s$ and one scattering angle, which is a simple function of the ratio $t/s$. Regardless of the parametrization, if we find an appropriate continuous variable $\alpha$ and function $F(\alpha)$ which appears erratic, we can analyze the spacings between zeros of $F(\alpha)$. If we can unfold the resulting ``spectrum'', or simply use the ratios $r_n$, then we should not be too sensitive to the choice of variable and be able to compare with the predictions of RMT. 

We can also generalize to a higher dimensional version where the scattering depends on several continuous variables, which we can denote generically as $\alpha^{(i)}$. In a similar way, one can define $F(z_n^{(i)}) = 0$ and analyze the spacing vectors $\delta^i_n = z^i_n - z^i_{n+1}$. In the following, we will focus only on the dependence on a single kinematical variable. For the four point scattering amplitude, as we will see, we will compute only chaotic behavior in the scattering angle $\theta$ when $s$ is large and fixed.

We now review the statistical distributions for the spacings and their ratios associated to random matrices.
 

\section{The \texorpdfstring{$\beta$}{beta}-ensemble of random matrices} \label{sec:beta}
\subsection{Distributions of eigenvalue spacings and their ratios}
The three classical Gaussian ensembles of random matrix theory  (RMT) \cite{Mehta:book,Akemann:book} are the Gaussian orthogonal ensemble (GOE), the unitary (GUE), and the symplectic (GSE). The eigenvalues of $N\times N$ matrices in these ensembles have the joint probability density function (PDF) which is
\be P_N(\lambda_1,\lambda_2,\ldots,\lambda_N) = {\cal C}(\beta) \times \exp\left(-\frac\beta2\sum_{i=1}^N \lambda_i^2\right) \prod_{1\leq i < j \leq N} |\lambda_i-\lambda_j|^\beta \label{eq:beta_lambda} \ee
where $\beta=1$, 2, or 4 for the GOE, GUE, and GSE respectively, and ${\cal C}(\beta)$ is a normalization constant. One can generalize from these three special cases by taking $\beta$ to be a continuous parameter, and defining in this way the $\beta$-ensemble for any $\beta>0$ starting from the above distribution.

One of the main objects of study in a given system are its level spacings. The distributions for the difference of eigenvalues $\delta = \lambda_2-\lambda_1$ of a $2\times 2$ matrix of the ensemble, properly normalized such that $\avg{\delta} = 1$, are given by the well known Wigner surmises:
\be p_{\beta}(\delta) = {\cal N}_{\beta} \delta^\beta \exp(-c_\beta \delta^2) \label{eq:beta_delta} \ee
with
\be {\cal N}_\beta = 2\frac{[\Gamma(\frac{\beta+2}2)]^{\beta+1}}{[\Gamma(\frac{\beta+1}2)]^{\beta+2}}\,,\qquad c_\beta = \left(\frac{\Gamma(\frac{\beta+2}2)}{\Gamma(\frac{\beta+1}2)}\right)^2\ee

Rather than the spacings themselves, we will focus on the ratios of consecutive spacings. In the minimal case of a $3\times3$ matrix with eigenvalues $\lambda_1 < \lambda_2 < \lambda_3$, the PDF of the ratio $r = (\lambda_3-\lambda_2)/(\lambda_2-\lambda_1)$ in the $\beta$-ensemble is \cite{Atas:2013dis}
\be f_\beta(r) = \frac{3^{\frac{3+3\beta}2}\Gamma(1+\frac\beta2)^2}{2\pi \Gamma(1+\beta)} \frac{(r+r^2)^\beta}{(1+r+r^2)^{1+\frac32\beta}} \label{eq:beta_r}\ee
The distribution is symmetric under $r\to \frac1r$ since
\be f_\beta(r) = \frac{1}{r^2}f_\beta(\frac{1}{r}) \ee
For larger $N\times N$ matrices it is shown numerically that the deviations from the distributions \eqref{eq:beta_delta} and \eqref{eq:beta_r} are small.

Equation \eqref{eq:beta_r} will be our main fitting model for the distribution of spacing ratios in string scattering amplitudes. We will also consider the variable $\tilde r \equiv \min(r,\frac1r)$. Thanks to the symmetry property of $r\to1/r$, the PDF of $\tilde r$ is simplify $2f_\beta(\tilde r)$ restricted to the range $\tilde r \in [0,1]$.

\subsection{Log-gases and the \texorpdfstring{$\beta$}{beta}-ensemble}
A physical system described by the $\beta$-ensemble is the log-gas \cite{Dyson:1962b,Forrester:book}, a Coulomb gas of $N$ charged particles interacting through a logarithmic potential between each pair,
\be V_{ij}(x_i,x_j) = \log|x_i-x_j| \ee
For our purposes, the system is taken to be one dimensional.

In addition to the pairwise interaction one should introduce a background charge density $\rho_0(x)$ such that the total charge is zero.  The neutrality condition is that $\int dx\rho_0(x) = -N$, since each particle is taken to have unit charge. The interaction of the background charge density with the $i$-th charge contributes a term
\be V_i(x_i) = \int dx^\prime \log|x-x^\prime| \rho_0(x^\prime) \ee
to the total potential.

Now, taking a charge density confined to the interval $(-\sqrt{2N},\sqrt{2N})$ and of the form
\be \rho_0(x) = -\frac{\sqrt{2N}}{\pi} \sqrt{1-\frac{x^2}{2N}} \ee
one finds that, up to a constant, $V_i(x) = -\frac{x^2}{2}$. The full potential is
\be V(x_1,\ldots,x_N) = \mathrm{Const.} - \sum_{i=1}^N \frac{x_i^2}{2}\, +\, \sum_{1\leq i < j \leq N}\log|x_i-x_j| \ee
Then, one can see that, for a given ``microstate'' corresponding to a choice of  $(x_1,x_2,\ldots,x_N)$, the Boltzmann factor $e^{-\beta V(x_1,\ldots,x_N)}$ takes exactly the form of eq. \eqref{eq:beta_lambda}, with $\beta$ precisely identified with the inverse temperature of the gas.\footnote{The Coulomb gas representation corresponding to a logarithmic potential finds many applications in 2D${-}$CFT including minimal models, Liouville models and many others \cite{DiFrancesco:Book,Ginsparg:1988ui,Cardy:2008jc}.}

\section{Highly excited string (HES) states} \label{sec:HES}
\subsection{Regge resonances and DDF operators} \label{sec:DDF}
The string spectrum contains an infinite number of massive higher-spin excitations often called Regge resonances. They lie on linear Regge trajectories in the plane $(J, M^2)$ with Regge slope $\alp$ and an intercept that depends on the model.For the open bosonic strings that we focus on the spectrum is given by
\be\label{HESgr1} \alp M^2 = N-1 \ee
where $N=\sum_{n} n g_n$ is the ``level'' and $J=\sum_{n=1} g_n\le N$ is the helicity of the string state, assuming only transverse oscillators are taken into account. 

The lowest lying states are easy to characterize. The ground state $N{=}0$ is a scalar tachyon with $J{=}0$ and $\alp M^2 {=}-1$. The first excited state $N{=}1$ is a massless vector boson ($J{=}1$, $\alp M^2 {=}0$). The first massive (second excited) level $N{=}2$ is a massive tensor boson made up of two terms, $J{=}2{=}g_1$ and $J{=}1{=}g_2$, such that $\alp M^2 {=}1 {=} (1+1)-1 {=} 2 - 1$. Note that the $J{=}1$ transverse ``vector'' polarizations are those needed to make the helicity $J{=}2$ states massive (the latter comprise a $J{=}0$ state related to the ``trace'' over the transverse indices, i.e. their contractions).

The situation soon becomes very messy since the degeneracy of the states with the level grows (for large $N$ and up to a constant) as
\be\label{HESgr2}
d_N \simeq N^{-{c{+}3\over 4}}\exp(2\pi \sqrt{{c\over 6} N}) =   N^{-{27\over 4}} \exp(4\pi \sqrt{N}) \quad {\rm for} \quad c=c_t=24=26-2
\ee
which counts the partitions of the integer $N$, encoded in Dedekind $\eta$ function
\be\label{HESgr3}
{\cal Z}_B(q) = {1\over \eta(q)^{24}} =  {1\over q \prod_{n=1}^\infty (1-q^n)^{24}} = 
{1\over q} (\sum_k d_k q^k)^{24}
\ee
The situation drastically simplifies for the first Regge trajectory with $J{=}N$, i.e. for the states with maximal spin at a given level. These correspond to $g_1{=}N$ and all the other $g_k{=}0$ for $k>1$. 

Counting states with a given spin $S$ at level $N$ is quite involved. One starts with the multi-helicity (super)trace
\be\label{HESgr4}
{\cal B}(q, \alpha_I) = {1\over q \prod_{I=1}^{12} \prod_{n=1}^\infty (1-e^{i\alpha_I} q^n) (1-e^{-i\alpha_I} q^n)} 
\ee
and then expands in characters of $SO(24)$. Setting $\alpha_I=\alpha$ for all $I=1,...12$ counts the total ``helicity'' $J$, which is less than the full actual ``spin'', $J\le S\le S_{Max}=N$, where $S$ is what classifies the representations of $SO(25)$. In fact tensors with mixed symmetry can appear for which the notion of spin is not even well defined.

If one focuses on 3 spatial directions the situation improves since one can put 
$\alpha_1=\alpha$ and all the rest to zero, $\alpha_I=0$ for $I=2,...12$. The relevant characters of $SU(2)\sim SO(3)\supset SO(2)_T$ are identical to the Gegenbauer polynomials of parameter $1$\footnote{These are relevant in the scattering in $D=5$ as in \cite{Bianchi:2017sds, Bianchi:2018kzy}.}
\be\label{HESgr5}
{\cal X}_J(\alpha) = \sum_{m=-J}^{m=J} e^{i \alpha m} = {\sin {(2J+1){\alpha\over2}} \over 
\sin {\alpha\over2}}
\ee
Using orthogonality of the characters 
\be\label{HESgr6}
{\cal B}(q, \alpha_1, \alpha_{I\neq 1}=0) = {1\over q} \sum_N d_N(J) q^N {\cal X}_J(\alpha)
\ee
one has
\be\label{HESgr7}
 d_N(J) = \oint {dq\over q^N} \int {d\alpha\over 2\pi} {\cal X}_J(\alpha) {\cal B}(q, \alpha_1, \alpha_{I\neq 1}=0) 
 \ee
Later on we will give an estimate of $d(N,J)$.

Let us consider other physical properties of HES states. Regge resonances are very narrow (zero width) for string coupling $g_s{=}0$ (free string) but acquire a finite width when $g_s\neq 0$. To lowest order in $g_s$, their decay amplitude corresponds to a 3-point function on the disk 
\be\label{HESgr8}
{\cal A}(p_1,p_2,p_3) = \langle V(p_1,H_1) V(p_2,H_2)V(p_2,H_2)\rangle
\ee
where $V$ is a BRST invariant vertex operator with momentum $p$ and ``polarization'' $H$. Let us focus on the decay amplitude of a massive higher-spin state at level $N$ into two tachyons.

While the BRST invariant vertex operator for a tachyon is simply 
\be\label{HESgr9}
V_T= e^{ipX}
\ee
with $p^2{=}-M^2{=} 1/\alp$, writing down the most general ``covariant'' vertex operator for massive states is very challenging due to the large ``gauge symmetry''. The problem can be overcome by relying on the time-honoured DDF (after Del Giudice, Di Vecchia, Fubini) approach \cite{DelGiudice:1971yjh}. 

The DDF construction is based on the choice of an arbitrary tachyonic momentum $p$ ($p^2{=} 1/\alp $) and a null momentum $q$ ($q^2{=}0$) chosen in such a way that $2\alp  p{\cdot}q{=} -1$. A massive on-shell momentum at level $N$ obtains 
\be\label{HESgr10}
p_N = p - Nq
\ee
that suggests a physical interpretation of the excited  string state as the tachyonic state with momentum $p$ that successively emits (absorbs) massless vector bosons with momentum $q$ (-$q$). With a choice of $q$ one can define the DDF operators
\be\label{HESgr11}
A_n^i(q) = \oint {dz\over {2\pi}} \partial X^i e^{in qX}
\ee
where $i$ runs over the transverse directions, i.e. $q{\cdot}A{=}0$. 
The most general BRST invariant state can be written as
\be\label{HESgr12}
\Big|\{g_n\}: N{=}\sum_n n g_n,\, p_N {=} p{-}Nq\Big\rangle = \prod_{n=1}^\infty A_{-n}^{i_n}(q) |0, p\rangle
\ee

\begin{figure}[t!]
\centering
\includegraphics[width=0.66\textwidth]{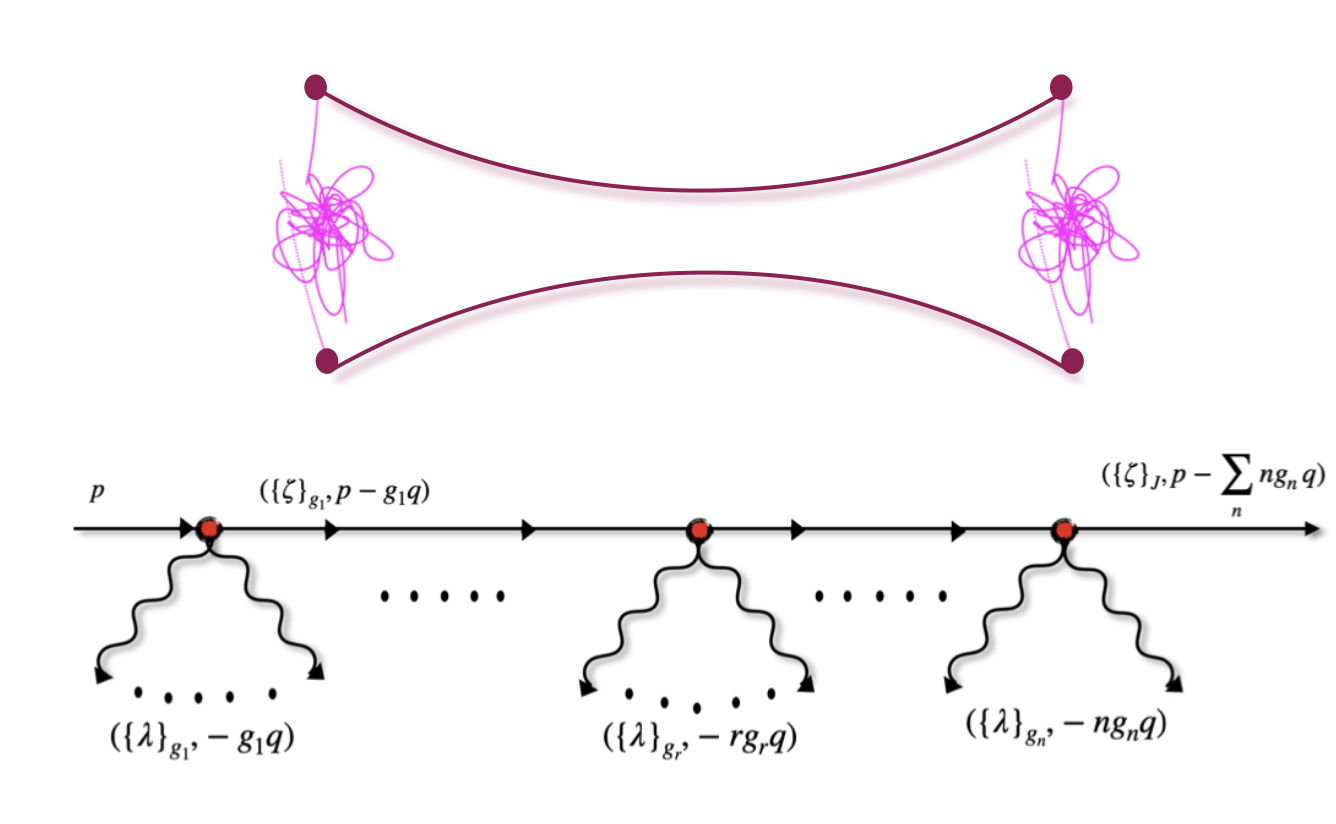}
\caption{Above: a representation of the profile of a generic highly excited open string state, where the magenta lines describe the complexity of the profile. Below: the DDF construction of such state. We start from a tachyonic vacuum state with momentum $p$ and we use the iterative action (codified in the operator product expansion represented by the red points) of DDF creation operators (which are described by the photon lines with polarizations $\{\lambda\}$ and momentum $-ng_{n}\,q$). We end up with a final state $H^{(i)}_N(\{\zeta\};p_{N} )$ with final momentum $p_{N}=p-q\sum_{n}ng_{n}$ and polarizations $\{\zeta\}$, as in eq. \eqref{HESgr13}.}
\end{figure}

For simplicity let us focus on the level $N$ states of the form
\be\label{HESgr13}
 \prod_{n=1}^\infty (\lambda{\cdot}A_{-n}(q))^{g_n} |0, p\rangle=\big|H^{(i)}_N(\{\zeta\})\big\rangle
\ee
where $\lambda$ is a complex transverse (``null'' in the sense that $\lambda{\cdot}\lambda = 0$) vector polarization and $\zeta=\lambda{-}2\alp (\lambda{\cdot}p) q$ such that $p{\cdot}\zeta=q{\cdot}\zeta = 0$. For this state we can compute for instance its decay amplitude into two tachyons with momenta $p_1$ and $p_2$ with $p{-}Nq + p_1+p_2=0$ for momentum conservation. 

Moreover if the initial excited string state had definite spin $\vec{S}$ (whichever the mass/level\footnote{Even superposition of mass eigenstates with different masses would give a 'trivial' angular distribution as long as all the spins are aligned ie $\vec{J}_N=\vec{J}_N'$ in the superposition.}) the amplitude would simply be
\be\label{HESgr14}
{\cal A}_{HTT}= C_{S}^{(H)} H_S{\cdot}(\vec{p}_1{-}\vec{p}_2)^{\otimes s} 
\ee
so that the angular distribution of the products (tachyons) would be completely fixed to be the Legendre polynomial $P_{S}(\vec{n}{\cdot}\vec{n}_p)$ with $\vec{n}_p={\vec{p}_1{-}\vec{p}_2\over |\vec{p}_1{-}\vec{p}_2|}$ in the rest frame of the decaying particle, whereby $\vec{p}_2=-\vec{p}_1$, while $|\vec{p}_1| = |\vec{p}_2| = {1\over \alp } {+} {M^2\over 4}$.

This shows no chaotic behavior, for instance this would be the case for states of the first Regge trajectory. 
On the contrary using the exponential degeneracy at level $N$ one can consider a generic state $H^{(i)}_N$ without definite helicity $J$ or even with definite helicity but without definite spin $S$.  This exposes chaotic behavior due to the random and highly erratic superposition of spin components in the generic state. As shown in \cite{Gross:2021gsj} the erratic behavior is exposed even for similar partitions.

\subsection{Integer partitions} \label{sec:partitions}
Since our HES states are characterized by a choice of a partition of the level $N \sim M^2$, we discuss here some basic properties of integer partitions that will be useful in the following.

The state $H_N^{(i)}$ is a state constructed from the DDF operators with a particular choice of polarizations,
\be\label{HESgr15} |H_N^{(i)}\rangle = \prod_{n=1}^N \left(\lambda \cdot A_n\right)^{g_n} |0\rangle \ee
and is defined by the integer partition \(\{g_n\}\) for which
\be\label{HESgr16} N = \sum_{n=1}^\infty n\, g_n\,, \qquad J = \sum_{n=1}^\infty g_n \ee
For fixed \(\lambda\), the number of states of this form at level \(N\) is equal to the total number of integer partitions of \(N\). In our notation $(i)$ is an index enumerating the states at level $N$ which needs no precise definition.

The total number of partitions of $N$ into at most $J$ summands can be computed from the generating function
\be\label{HESgr17} F_J(x) = \sum_{N=1}^\infty C_{N,J} x^N = \prod_{k=1}^J \frac{1}{1-x^k} \ee
and therefore the number of partitions of length exactly $J$ is given by
\be\label{HESgr18} p_{N,J} = C_{N,J} - C_{N,J-1} \ee
The total number of partitions of $N$ is also given by the above generating function as $p_N {=} C_{N,N}$. The number of partitions behaves for large \(N\) as
\be\label{HESgr19} p_N \approx \frac{1}{4\sqrt{3} N} e^{C \sqrt{N}}\,, \qquad C \equiv \pi \sqrt{\frac23} \ee
The number of partitions of fixed length $J$ is asymptotically given by a Gumbel distribution \cite{Erdos:1941}, i.e. according to the probability density function
\be d_N(J) = \frac{1}{{\nu}} \exp\left(-\frac{J-\mu}{{\nu}} - e^{-\frac{J-\mu}{{\nu}}}\right) \label{eq:Gumbel} \ee
with the parameters scaling as \(\mu \sim \sqrt{N} \log N\) and \({\nu} \sim \sqrt{N}\). More specifically the distribution is centered around
\be \langle J \rangle \approx \frac{1}{C} \sqrt{N}\log N \label{eq:typicalJ} \ee
which makes this the ``typical helicity'' of a randomly chosen string state at level \(N\).

In addition to $J$ another useful parameter of a given partition is $n_{\text{max}}$, the largest summand in the partition, i.e. the largest $n$ for which $g_n \neq 0$. It is easy to see that $n_{\text{max}}$ is exactly the length of the conjugate partition. The conjugation of integer partition is most easily understood as a rotation and reflection of the Ferrers diagram associated with the partition, exchanging rows for columns. For example
\[
\begin{tabular}{lcl}
$4+3+2+2+1$ &  & $5+4+2+1$ \\
$\bq\bq\bq\bq$  & & $\bq\bq\bq\bq\bq$ \\
$\bq\bq\bq$  & $\longleftrightarrow$ & $\bq\bq\bq\bq$ \\
$\bq\,\bq$  &  & $\bq\,\bq$ \\
$\bq\,\bq$  & & $\bq$ \\
$\bq$  & &  \\
\end{tabular}
\]
are two partitions of 12, conjugate to each other. Since conjugation is a one-to-one operation, it follows immediately that the number of partitions of $N$ into integers less or equal to $n$ is the same as the number of partitions of length $J=n$. This is particularly useful for generating random partitions of fixed length, as described in appendix \ref{app:randompartitions}.

\section{Chaos in the decay amplitude of a highly excited string} \label{sec:3pt}
\subsection{The decay amplitude}
We study the decay amplitude of a highly excited string state into two tachyons. The relevant kinematical variable is $\alpha$, the angle between the outgoing tachyons and the photons used to create the DDF state, i.e. the angle between the momentum of one of the tachyons and the photon momentum $q$ seen e.g. in eq. \eqref{HESgr13}.

The angular dependence of the full amplitude can be compactly written as \cite{Gross:2021gsj}
\be {\cal A}_{H_N^{(i)}\to TT} \propto (\sin \alpha)^J  \prod_{n=1}^{N} \left(\sin(\pi n \cos^2\frac\alpha2)\frac{\Gamma(n \cos^2 \frac\alpha2)\Gamma(n \sin^2 \frac\alpha2)}{\Gamma(n)}\right)^{g_n} \label{eq:A_HTT} \ee
where $H_N^{(i)}$ is a state defined by an integer partition $\{g_n\}$ of level $N$ and with helicity $J$, as described in the previous section.

Another representation of the amplitude can be reached using basic identities of the Gamma function, $\Gamma(z+1) = z \Gamma(z)$ and $\Gamma(z)\Gamma(1-z)\sin(\pi z) = \pi$. Then, each term in the product is a a polynomial in $x \equiv \cos^2\frac\alpha2$ of degree $m-1$, which can be written compactly using the Pochhammer symbol,
\begin{align}
    \frac{\Gamma\big(n(1-x)\big)\Gamma(n x) \sin(n\pi x)}{\Gamma(n)} =&\,\frac{\pi}{(n-1)!}(1-nx)_{n-1} \nonumber \\
      =&\,\pi (1- \frac{n x}{n-1})(1- \frac{nx}{n-2})\ldots(1- \frac{nx}{1})
\end{align}
so that the amplitude is written most compactly as
\be {\cal A}_{H_N^{(i)}\to TT} \propto (\sin \alpha)^J  \prod_{n=1}^{N}\left(\frac{(1-n\cos^2\frac\alpha2)_{n-1}}{\Gamma(n)}\right)^{g_n} \label{eq:A_HTT_poly} \ee

It is convenient for computation and visualization to work with the logarithmic derivative,
\be F(\alpha) \equiv \frac{d}{d\alpha}\log {\cal A}  = J \cot \alpha - \frac12\sin\alpha \sum_{n=1}^{N} g_n\sum_{k=1}^{n-1}\frac{n}{n-k-n\cos^2\frac\alpha2}   \ee
The points $z_n$ at which $F(z_n)=0$ are the local maxima and minima of ${\cal A}(\alpha)$, and they are always maxima of the absolute value $|{\cal A}|$. For this reason we can refer to all the extremal points of the amplitude collectively as its peaks. 

In terms of $x = \cos^2\frac\alpha2$, the peaks are located at the points where
\be F(x) = \frac{1}{2\sqrt{x(1-x)}}\left(J (2x-1) - 2 x(1-x)\sum_{n=1}^{N} g_n\sum_{k=1}^{n-1}\frac{n}{k-nx}\right) = 0 \ee 
which is a polynomial equation in $x$.

Our object of study is the distribution of the ratios of the spacings between consecutive solutions of $F(\alpha)=0$.




\subsection{Statistical analysis of spacing ratios}
Our main result is that the ratios $r_n$ of the spacings between consecutive peaks of the amplitude \eqref{eq:A_HTT} are distributed as predicted by the $\beta$-ensemble, i.e. as in the distribution of eq. \eqref{eq:beta_r}, with $\beta$ depending on the parameters of the HES state. In the following we describe in detail the statistical analysis that leads to this result.

\subsubsection{Fitting model and selection of states} \label{sec:fitmodel}
For a generic state at level $N$, the number of peaks in the decay amplitude (or zeros of its logarithmic derivative) scales linearly with $N$. For very large values of $N$, say $N \sim $ 10,000, there would be sufficient zeros such that one could measure the distribution of spacing ratios in a single amplitude of a particular excited string state. For intermediate $N$, meaning of order 100, we collect data from many different states in order to perform the statistical analysis.

Denoting the set of ratios for a specific state $H_N^{(i)}$ as $\{r_n\}_{N^{(i)}}$, we will study the distribution of the values in the union of many such sets, which we can write as
\be \{r_n\}({\cal S}) \equiv \bigcup_{N^{(i)}\in {\cal S}} \{r_n\}_{N^{(i)}} \ee
for some specific sample subset of states ${\cal S}$.

Once we have a set of $\{r_n\}$ with enough data points, we fit their distribution to that of $r$ in the $\beta$-ensemble of eq. \eqref{eq:beta_r}. There is a single continuous and positive fitting parameter, $\beta$, which can be directly related to the average value of the distribution $\avg{r}$. We will examine the behavior of the average $\avg{r_n}$ in detail in the following.

Since the space of all possible states is too large to study, we must choose wisely which states would go into a representative sample. We observe from the data that there is a dependence of $\avg{r_n}$ on the level number $N$ and the helicity $J$. It is then assumed, as a model, that the fitting parameters depend only on $N$ and $J$, though the full physical picture could be more complex.

There are some technical difficulties in selecting random states to go into a sample. Our aim is to choose ``generic'' states once we have fixed $N$ and (optionally) $J$. The most straightforward way is to select at random a small sample from the full list of possible states, with each state having an equal probability of being chosen. At some point this becomes impractical as the list of possible states grows exponentially. It is non-trivial to devise an algorithm that will generate random partitions of a large integer in such a way that all partitions are equally likely to be chosen (see appendix \ref{app:randompartitions} for details). For our purposes it is important not to introduce any biases in the selection of states, since we cannot predict how that might affect the results. Once we fix the parameters, the selection will be random, with all partitions of the given $N$ (and $J$ when that is fixed) having equal probability to go into the sample. Our samples will typically consist of thousands of states, which will provide more than enough data to observe a distribution of the spacing ratios, but we keep in mind that the states in the sample still represent only an exponentially small fraction of the states of the highly excited string.

\subsubsection{Results}
The results of our analysis indicate that the distribution of spacings of peaks of the amplitude is well modeled by the random matrix formula of the $\beta$-ensemble, with the $\beta$ parameter depending on the decaying state. Specifically we measure the dependence on the level $N$ and the helicity $J$. The results are collected in tables \ref{tab:HTTnj}--\ref{tab:HTTj}.

\begin{figure}[ht!]
    \centering
    \includegraphics[width=0.40\textwidth]{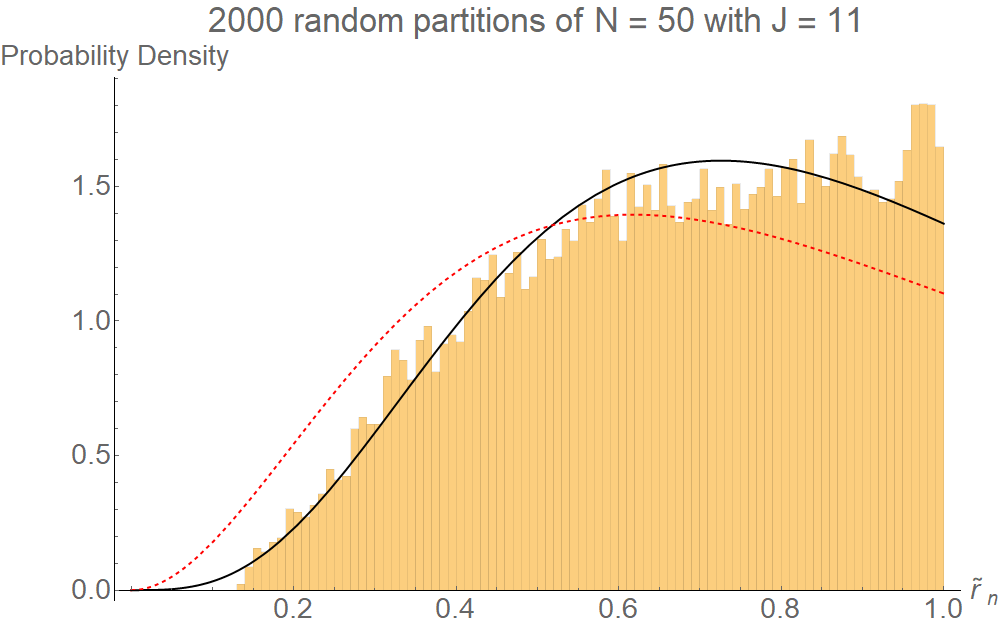}
    \includegraphics[width=0.40\textwidth]{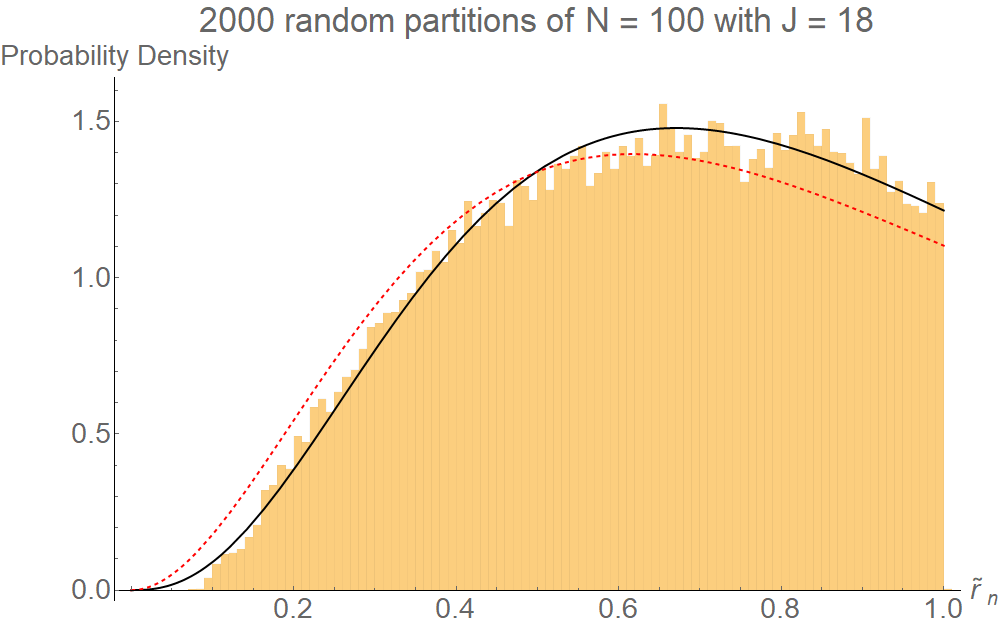}\\
    \includegraphics[width=0.40\textwidth]{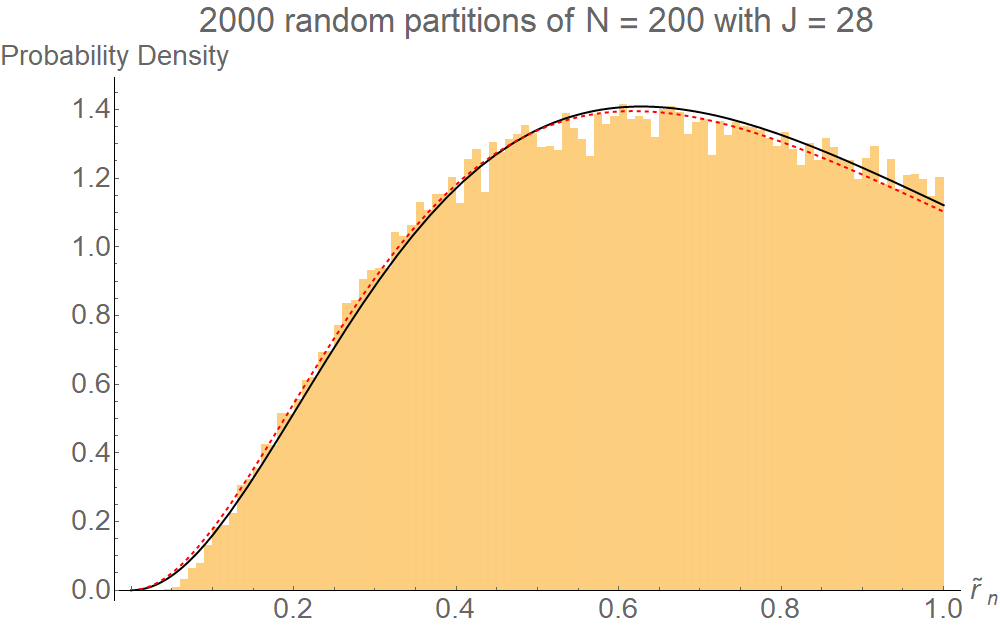}
    \includegraphics[width=0.40\textwidth]{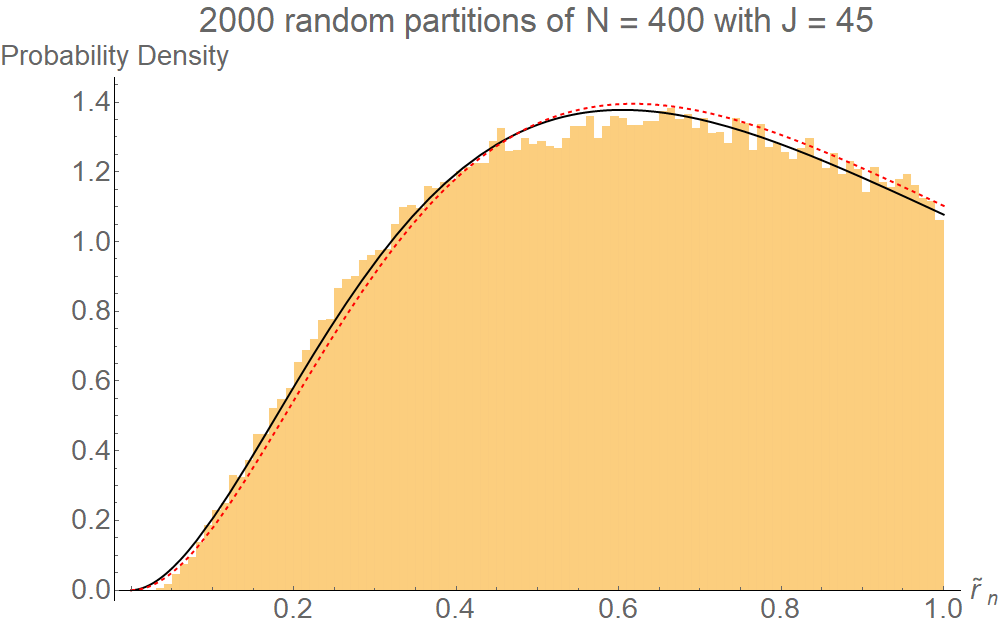}\\
    \includegraphics[width=0.40\textwidth]{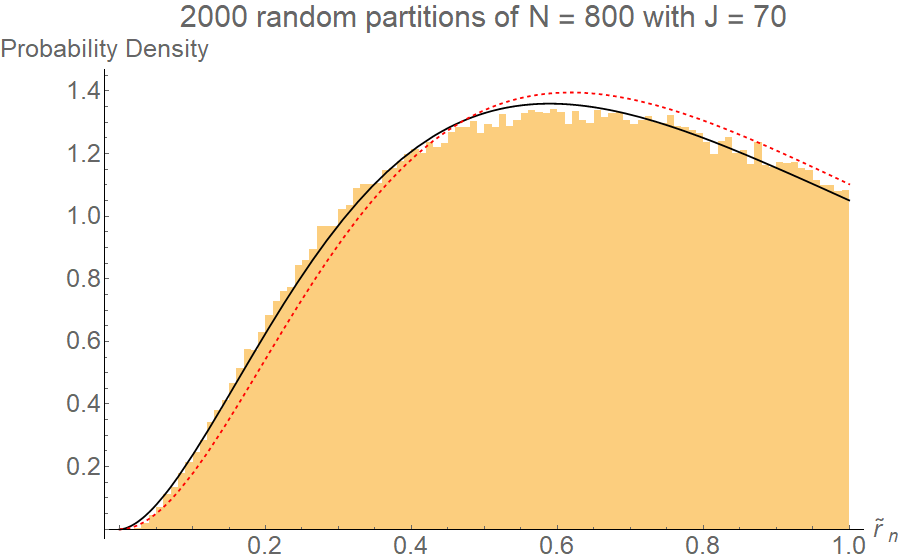}
    \includegraphics[width=0.40\textwidth]{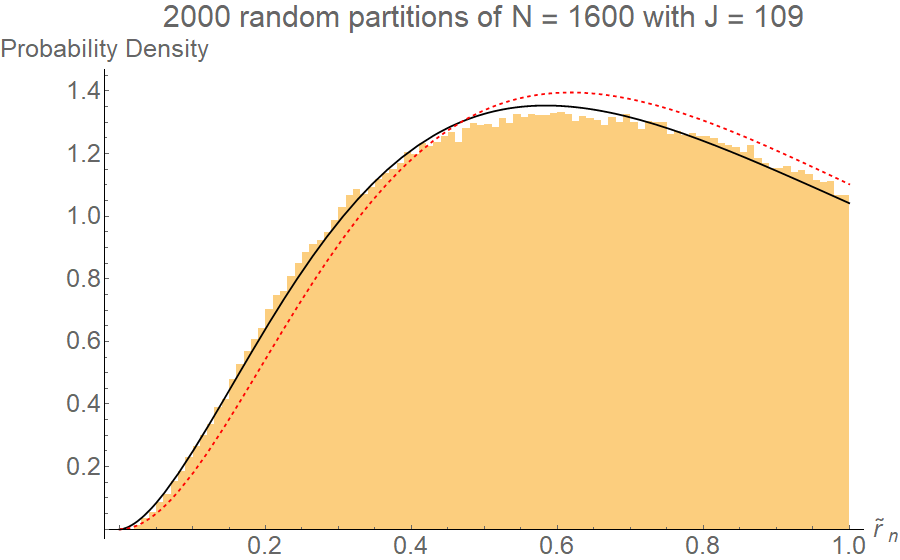}\\
    \caption{The measured distributions of $\tilde r$ across samples of 2000 states chosen at random with fixed $N$ and $J$, fitted to a $\beta$-ensemble distribution (black line). The distribution is also compared to the GUE distribution (red, dashed). The fitting parameters are listed in table \ref{tab:HTTnj}.}
    \label{fig:r_HTT}
\end{figure}

In the present case we have analyzed states up to $N=1600$. The results are generally well fitted by the predicted distribution. There are some deviations, but these become small as we increase $N$, see figure \ref{fig:r_HTT}.

We observe a monotonous increase of the average $\avg{r_n}$ with the level $N$. The growth appears to be logarithmic at most, and it is impossible to tell from the data whether it continues indefinitely as $N \to \infty$ or approaches some asymptotic value. In \cite{Bianchi:2022mhs} we noted that there is a similar dependence of $\avg{r_n}$ when examining the spacings of the non-trivial zeros of the Riemann zeta function. There, the average increased slowly as one covered a larger and larger range of zeros, but it eventually approaches the value predicted from the GUE. The numerical evidence for this is very strong, thanks to the vast range of zeros of the zeta function that has been computed \cite{Odlyzko:Zeta,Odlyzko:2001}.

For reference, the predicted averages of $\langle r \rangle$ ($\langle \tilde r \rangle$) are, $1.75$ ($0.536$) for GOE, $1.361$ ($0.603$) for GUE. As a function of $\beta$, $\langle r\rangle$ is a monotonously decreasing function. It diverges at $\beta=0$ and approaches 1 at large $\beta$. The average of $\langle \tilde r$ increases from $\approx0.408$ at $\beta=0$ to 1 at large $\beta$.\footnote{The distribution at $\beta=0$ is close to but not the same as the Poisson distribution, for which $\langle r\rangle $ diverges while $\langle \tilde r \rangle \approx 0.386$.}

A question that one can ask is whether spacing ratios in the string amplitudes also approach the GUE distribution at very large $N$. From table \ref{tab:HTTnj} one can see that around $N=200$ the best fit for $\beta$ is close to the GUE value of 2, but as we increase $N$ we reach $\beta<2$. The deviation that we find at $N=1600$, the largest value examined, is big enough to suggest that $2$ is not a good asymptotic value for $\beta$ at large $N$. The GOE value of $\beta=1$ is still theoretically possible, but at the presently examined values of $N$ we are still far from it, and the rate at which $\beta$ decreases is already slowed down. A simple fit of the form $\beta(N) = a + b/N$ is in good agreement with all the values quoted in table \ref{tab:HTTnj} and suggests $\beta = 1.68$ as the asymptotic value. On the other hand, a continuing logarithmic decrease cannot be ruled out from the data.

\begin{figure}[ht!]
    \centering
    \includegraphics[width=0.60\textwidth]{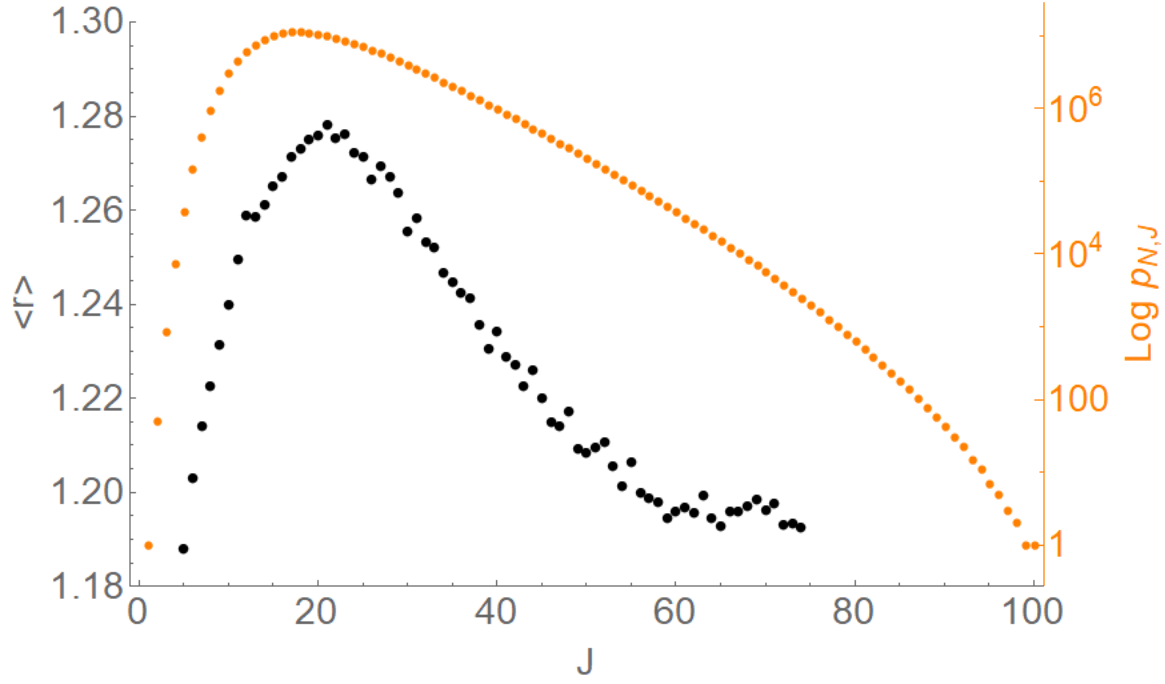}
    \caption{The average value $\avg{r_n}$ as a function of $J$, where $N$ is fixed at 100. The black dots are the measured values of $\avg{r_n}$ on samples of 2000 partitions at each helicity $J$. Also drawn is the number of states of a given $J$ at $N=100$, on a logarithmic scale. Some specific values are in table \ref{tab:HTTj}.}
    \label{fig:rofJ}
\end{figure}

In addition, there is a dependence on $J$ when $N$ is fixed. In this case the average is greater for values of $J$ for which there is a larger degeneracy. This is visible in figure \ref{fig:rofJ}, where we plot $\avg{r_n}$ alongside the logarithm of the number of partitions of length $J$ at level $N=100$. The plots do not match exactly, in particular the maximum points are at different values of $J$, but a correlation is visible. Then, the common observation from the dependence on $N$ and $J$, is that the average appears to grow in correlation with the number of degenerate states at the given level, but in a slow, at most logarithmic fashion.

\begin{table}[ht!]
    \centering
    \begin{tabular}{|c|c|c|c|c|c|c|c|} \hline
         $N$ & $J$  &   Total number & Points & Per & Average & Average & Fitted \\
          & &        of states                  & in sample    & state     & \(\langle r_n \rangle\) & \(\langle \tilde r_n \rangle\) & $\beta$\\ \hline\hline

        50 & 11   & 17,475     & 46,354 & 24 & 1.206 & 0.662 & 3.36  \\ \hline
               
        75 & 15   & 552,767     & 69,247 & 34 & 1.247 & 0.641 & 2.81 \\ \hline

          100 & 18 & \(11.1\times 10^6\)    & 92,251 & 46 & 1.271 & 0.628 & 2.55  \\ \hline

            150 & 23 &  \(1.90\times 10^9\)    & 139,428 & 70 & 1.307 & 0.614 & 2.26 \\ \hline
            
            200 & 28 &  \(158\times 10^9\)   & 184,705 & 90 & 1.333 & 0.606 & 2.09 \\ \hline

            300 & 37 & \(295\times 10^{12}\)   & 276,244 & 138 & 1.357 & 0.599 & 1.96 \\ \hline

            400 & 45 & \(184\times 10^{15}\) & 370,123 & 186 & 1.372 & 0.596 & 1.88 \\ \hline

            800 & 70 & \(1.08\times10^{26}\) & 728,048 & 362 & 1.400 & 0.590 & 1.76 \\ \hline

            1600 & 109 & \(4.22\times10^{38}\) & 1,446,008 & 720 & 1.413 & 0.588 & 1.72 \\ \hline
         
    \end{tabular}
    \caption{Dependence of $\avg{r}$ and $\beta$ on $N$, based on samples of 2000 states at each $N$ and $J$. The value of $J$ is fixed to the value with a maximum number of partitions. The distributions are plotted in figure \ref{fig:r_HTT}.}
    \label{tab:HTTnj}
\end{table}

\begin{table}[ht!]
    \centering
    \begin{tabular}{|c|c|c|c|c|c|c|} \hline
         $N$ &   Total number & Points & Per & Average & Average & Fitted \\
          &        of states                  & in sample    & state     & \(\langle r_n \rangle\) & \(\langle \tilde r_n \rangle\) & $\beta$\\ \hline\hline

          50 & 204,226 & 215,980 & 22 & 1.194 & 0.670 & 3.58  \\ \hline

          60 & 966,467 & 261,619 & 26 & 1.213 & 0.660 & 3.27  \\ \hline

          80 & \(15.8\times10^6\) & 352,526 & 34 & 1.244 & 0.644 & 2.87 \\ \hline

        100   & \(191\times 10^6\)  & 441,100 & 44 & 1.266 & 0.633 & 2.62 \\ \hline

        150   & \(40.9\times 10^9\)   & 668,831 & 66 & 1.301 & 0.618 & 2.32 \\ \hline

        200   & \(3.97\times 10^{12}\) & 886,007  & 88 & 1.325 & 0.610 & 2.15  \\ \hline
         
    \end{tabular}
    \caption{Dependence of $\avg{r}$ and $\beta$ on $N$, based on samples of 10,000 random partitions of $N$. The value of $J$ is not fixed, but distributed as predicted by \eqref{eq:Gumbel}.}
    \label{tab:HTTn}
\end{table}

\begin{table}[ht!]
    \centering
    \begin{tabular}{|c|c|c|c|c|c|} \hline
         $J$ &   Total number & Points & Per & Average & Fitted \\
          &        of states                  & in sample    & state     & \(\langle r_n \rangle\) & $\beta$\\ \hline\hline

           6 & 143,247 & 155,162 & 80 & 1.203 & 3.60 \\ \hline
         
        10  &\(2.98\times 10^6\)  & 126,008 & 64 & 1.241 & 2.95 \\ \hline

           14 & \(8.86\times 10^6\)   & 105,502 & 54 & 1.263 & 2.65  \\ \hline

            \bf{18} & \(11.1\times 10^6\)   & 92,251 & 46 & 1.271 & 2.55  \\ \hline

             22 & \(9.24\times 10^6\)   & 83,405 & 42 & 1.276 & 2.52  \\ \hline
           
             26 & \(6.32\times 10^6\)   & 76,211 & 38 & 1.272 & 2.57  \\ \hline

             30 & \(3.91\times 10^6\)    & 70,650 & 30 & 1.262 & 2.69  \\ \hline

             50 & 204,226    & 51,287 & 26 & 1.209 & 3.38  \\ \hline

             70 & 5604   & 31,060 & 16 & 1.197 & 3.50  \\ \hline

    \end{tabular}
    \caption{Dependence of $\avg{r}$ and $\beta$ on $J$ for $N=100$. For each $J$ we take 2000 states chosen at random. See plot in figure \ref{fig:rofJ}.}
    \label{tab:HTTj}
\end{table}

\section{Chaotic four point scattering process: one HES and three tachyons} \label{sec:4pt}
\subsection{The scattering amplitude}
The aim of this section is to analyze the chaotic behavior of the simplest four-point scattering amplitude involving one generic HES, {\it i.e.} scattering of three tachyons and one HES. In particular we will present the construction of the most general scattering amplitude based on the combination of the DDF formalism and string coherent state formalism, which is a suitable tool for the building of generating functions of amplitudes \cite{Bianchi:2019ywd}.
\begin{figure}[h!]
\centering
\includegraphics[width=0.48\textwidth]{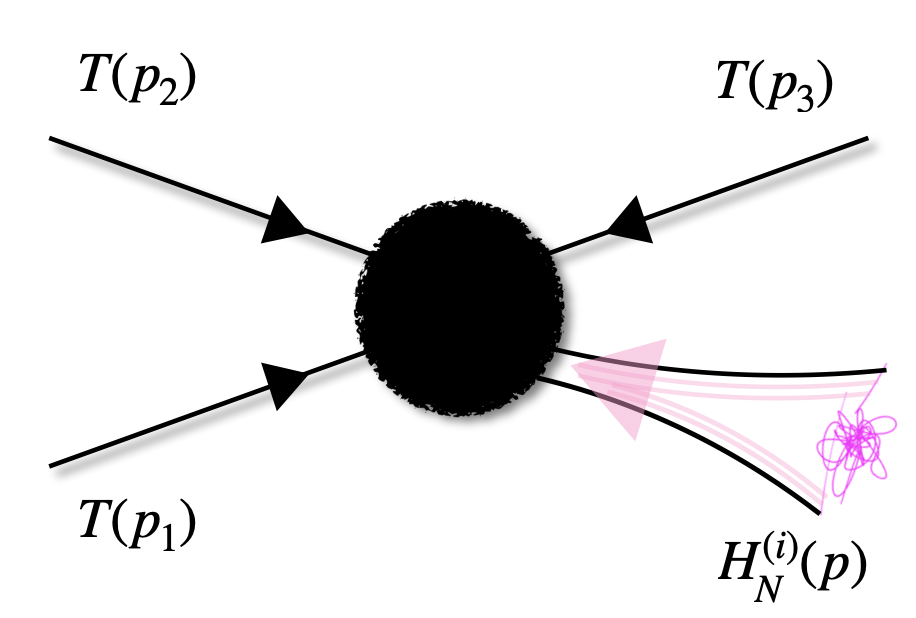}
\caption{Representative picture of the process under analysis.}
\label{pic4pt}
\end{figure}

The cross symmetric generating scattering amplitude in figure \ref{pic4pt} is composed of the following amplitudes\footnote{We set $2\alp=1$.} 
\be\label{scat1}
{\cal A}_{gen}^{HES}(s,t,u)={\cal A}_{gen}^{HES}(s,t)+{\cal A}_{gen}^{HES}(t,u)+{\cal A}_{gen}^{HES}(s,u)
\ee
where
\be\label{scat2}
{\cal A}_{gen}^{HES}(t,u)={\cal A}_{gen}^{HES}(s,t)\Big|_{p_{2}\Leftrightarrow p_{3}}\,,\quad {\cal A}_{gen}^{HES}(s,u)={\cal A}_{gen}^{HES}(s,t)\Big|_{p_{1}\Leftrightarrow p_{2}}
\ee

The generating amplitude of the process is given by
\be\label{ampint}
 {\cal A}_{gen}^{HES}(s,t)=\int_{0}^{1}dz \, z^{-{s\over2} -2} (1-z)^{-{t\over 2}-2} \,e^{ \sum_{n}{\cal J}_{n}{\cal O}_{n}(z) +\sum_{n,m}{\cal J}_{n}{\cal J}_{m}{\cal M}_{n,m}(z)}
\ee
where the contribution proportional to the contractions $\zeta_{n}{\cdot}p_{j}$ is given by
\be\label{es1}
{\cal O}_{n}(z)={\cal T}^{(2)}_{n}(q{\cdot}p_{3},q{\cdot}p_{2};z) + {\cal T}^{(3)}_{n}(q{\cdot}p_{3},q{\cdot}p_{2};z)
\ee
with the explicit form of the single contributions proportional to $\zeta_{n}{\cdot}p_{2}$ and $\zeta_{n}{\cdot}p_{3}$ (app. B)
 \be
{\cal T}^{(2)}_{n}(q{\cdot}p_{3},q{\cdot}p_{2};z)=z\, \zeta_{n}{\cdot}p_{2} {(nq{\cdot}p_{3})_{n{-}1}\over \Gamma(n)}  \, \pFq{2}{1}{ 1+nq{\cdot}p_{2}, 1{-}n}{ 2{-}n(1{+}q{\cdot}p_{3})}{ \,z} \label{eq:T2gen}
\ee
\be
 {\cal T}^{(3)}_{n}(q{\cdot}p_{3},q{\cdot}p_{2};z)=\zeta_{n}{\cdot}p_{3}  {(1{+}nq{\cdot}p_{3})_{n{-}1}\over \Gamma(n)}\,\, \pFq{2}{1}{ nq{\cdot}p_{2}, 1{-}n}{ 1{-}n(1{+}q{\cdot}p_{3})}{ \,z} \label{eq:T3gen}
\ee
while the contribution proportional to the contractions $\zeta_{n}{\cdot}\zeta_{m}$ can be written as 
\be
\begin{split}
{\cal M}_{n,m}(z)=\zeta_{n}{\cdot}\zeta_{m}\sum_{k=1}^{m}k\,&{(nq{\cdot}p_{3})_{n{+}k}\over (n{+}k)!}\, \pFq{2}{1}{ nq{\cdot}p_{2}, -n{-}k}{ 1{-}n(1{+}q{\cdot}p_{3}){-}k}{ \,z}\\
&{(mq{\cdot}p_{3})_{m{-}k}\over(m{-}k)!}\, \pFq{2}{1}{ mq{\cdot}p_{2}, -m{+}k}{ 1{-}m(1{+}q{\cdot}p_{3}){+}k}{ \,z} \label{eq:Mnmgen}
\end{split}\ee
The projection of the amplitude onto a precise HES state insertion can be realized as follows
\be\label{scat3}
{\cal A}(T(p_{1}),T(p_{2}),T(p_{3}),H_{N}(\{g_{\ell}\};q,p))=
\prod_{\ell=1}^{N}\left({d \over d{\cal J}_{\ell}} \right)^{g_{\ell}} {\cal A}_{gen}^{HES}(s,t)\Big|_{\{{\cal J}_{n}\}=0}
\ee
since the identification between a precise HES state and the corresponding amplitude can be seen through the relation
\be\label{scat4}
\prod_{\ell=1}^{N}{d^{g_{\ell}}\over d{\cal J}^{g_{\ell}}_{\ell}} \,e^{\sum_{n}{\cal J}_{n}\lambda_{n}{\cdot}A_{-n}}|\widetilde{p}\rangle\Big|_{\{{\cal J}{=}0\}}=\prod_{\ell=1}^{N}\left(\lambda_{\ell}{\cdot}A_{-\ell}\right)^{g_{\ell}}|\widetilde{p} \rangle=|HES,\{g_{\ell}\}\rangle_{N}
\ee
where $\widetilde{p}$ is the tachyonic reference momentum of the DDF formalism.
 
A compact representation of the generating scattering amplitude can be given by promoting the coefficients ${\cal O}_{n}$ and ${\cal M}_{n,m}$ to derivative operators acting on a generating function of powers of $z$. In particular after the integration over $z$ one has
\be\label{ampgen}
{\cal A}_{gen}^{HES}(s,t)={\cal A}_{Ven}(s,t)e^{\sum_{n}{\cal J}_{n}{\cal O}_{n}\left({d\over d\beta}\right)+\sum_{n,m}{\cal J}_{n}{\cal J}_{m}{\cal M}_{n,m}\left({d\over d\beta}\right)}\pFq{1}{1}{ -\alp s{-}1}{ -\alp s{-}\alp t{-}2}{ \,\beta} \Bigg|_{\beta=0}
\ee 
this representation is a realization of the identification 
\be\label{scat5}
\sum_{\ell} C_{\ell}\int_{0}^{1} z^{\ell}\, z^{-\alp s-2}(1{-}z)^{-\alp t-2} = {\cal B}\left(-\alp s{-}1,\,-\alp t{-}1 \right)\sum_{\ell}C_{\ell} {(-\alp s{-}1)_{\ell}\over (-\alp s{-}\alp t{-}2)_{\ell}}
\ee
where the coefficients ${C}_{\ell}$ are dictated by combinations of (\ref{es1}) and (\ref{eq:Mnmgen}).
We can observe a complete factorization between the partition structure and the pole structure, leading to a dressing factor of the Veneziano amplitude
\be\label{scat6}
{\cal A}_{Ven}(s,t)={\cal B}\left(-\alp s{-}1,\,-\alp t{-}1 \right)
\ee

\subsection{The HES amplitude in the high energy fixed angle regime}
In the kinematical regime where $s,|t|\gg1$ and $s/t$ is fixed, one can study the behavior the amplitude looking at the saddle point analysis of (\ref{ampint}), where the saddle is located at $z^{*}=s/(s{+}t)$. Alternatively one can obtain the same result studying (\ref{ampgen}) in the fixed angle limit where the action of the derivative operators turns out to be reproduced by the replacement $d/d\beta\rightarrow s/(s{+}t)$. The amplitude in fixed angle regime is then
\be\label{ampgenFa}
{\cal A}_{gen}^{HES}(s,t)\Big|_{f.a}={\cal A}^{f.a}_{Ven}(s,t)\,e^{\sum_{n}{\cal J}_{n}{\cal O}_{n}\left({s\over s{+}t}\right)+\sum_{n,m}{\cal J}_{n}{\cal J}_{m}{\cal M}_{n,m}\left({s\over s+t}\right)}
\ee 
where the standard behavior of the Veneziano amplitude in this limit,
\be\label{scat6b}
{\cal A}^{f.a}_{Ven}(s,t) \sim e^{-s\log{s} -t\log{t}+(s{+}t)\log{(s{+}t)}}
\ee
gets combined with an exponential dressing factor obtained by evaluating \eqref{eq:T2gen}-\eqref{eq:Mnmgen} at the saddle point $z^{*}{=}s/(s+t)$.
To write the dressing factor more compactly, let us denote
\be  \rho(s,\theta) \equiv -q{\cdot}p_2 \ee
Following the kinematics (appendix \ref{app:4ptkin}), it always holds that $1{+}q{\cdot}p_3 {=} 2\rho$. Using this notation, the dressing factor will be made up of
 \be\label{scat7}
{\cal T}^{(2)}_{n}|_{z=z^{*}}=\frac{\zeta_n{\cdot}p_{2}}{\Gamma(n)} (2n\rho{-}n)_{n{-}1}  z^{*}\,\, \pFq{2}{1}{ 1{-}n \rho, 1{-}n}{ 2{-}2 n \rho}{z^{*}}
\ee
\be\label{scat8}
 {\cal T}^{(3)}_{n}|_{z=z^{*}}= \frac{\zeta_n{\cdot}p_3}{\Gamma(n)} (2n\rho{-}n{+}1)_{n{-}1} \,\, \pFq{2}{1}{-n\rho, 1{-}n}{ 1{-}2n\rho}{z^{*}} 
\ee
and
\be\label{scat8b}
\begin{split}
{\cal M}_{n,m}|_{z=z^{*}}=\zeta_{n}{\cdot}\zeta_{m}\sum_{k=1}^{m}k\,&{(-n{-}2n\rho)_{n{+}k}\over (n{+}k)!}\, \pFq{2}{1}{-n\rho, {-}n{-}k}{ 1{-}2n \rho{-}k}{z^{*}}\\
&{(-m{-}2m\rho)_{m{-}k}\over(m{-}k)!}\, \pFq{2}{1}{ -m \rho, {-}m{+}k}{1{-}2m\rho{+}k}{z^{*}}
\end{split}
\ee
The hypergeometric function ${}_2 F_1(a,1-n,c;z)$ for positive integer $n$ is a polynomial in $z$ of degree $n{-}1$. Using its explicit form one can rewrite the expressions as
\be\label{scat9}
{\cal T}^{(2)}_{n}|_{z=z^{*}}= \frac{\zeta_n{\cdot}p_2}{\Gamma(n)}\sum_{k=0}^{n{-}1}\binom{n{-}1}{k}(2n\rho{-}n)_{n{-}k{-}1} (1 {-}n \rho)_{k}\, (z^{*})^{k+1} 
\ee
\be\label{scat10}
{\cal T}^{(3)}_{n}|_{z=z^{*}}= \frac{\zeta_n{\cdot}p_3}{\Gamma(n)} \sum_{k=0}^{n-1}\binom{n{-}1}{k}(2n \rho{-}n {+} 1)_{n{-}k{-}1} (-n \rho)_{k}  \,(z^{*})^k
\ee
from which one can see the dependence on the functions $z^*$ and $\rho$ is a polynomial one. One can also write the polynomial form of ${\cal M}_{n,m}$ in terms of a similar expansion.

To simplify the analysis of the amplitude, we can go to the specific case of identical circular polarizations, namely taking $\zeta_n = \zeta$ for all $n$, with $\zeta^2=0$.

In the case of circular polarizations, the terms ${\cal M}_{n,m}$ drop out and the amplitude in the fixed angle regime can be written as 
\be\label{scat11}
{\cal A}^{HES}_{f.a} = {\cal A}_{Ven}^{f.a}(s,\theta) \prod_{n=1}^N\left({\cal T}^{(2)}_{n}\left(s,\theta\right)+{\cal T}^{(3)}_{n}\left(s,\theta\right) \right)^{g_{n}}
\ee

From the kinematics in appendix \ref{app:4ptkin}, specialized to the high energy, fixed angle regime one can use approximate forms for $\rho$, $z^{*}$, and $\zeta{\cdot}p_i$ when $s\gg 2N$, and reduce the expressions to
\be\label{scat12}
{\cal T}^{(2)}_{n}\left(s,\theta\right)= {z^{*}(\theta)\sqrt{s}\over 2 \Gamma(n)} {\big(\rho(\theta){\cos\theta }{-}1\big){\left(2n\,\rho(\theta){-}n\right)_{n{-}1}}}  \, \pFq{2}{1}{ {1{-}n\,\rho(\theta)}, 1{-}n}{ 2(1{-}n\,\rho(\theta))}{ \,z^{*}(\theta)}
\ee
\be\label{scat13}
{\cal T}^{(3)}_{n}\left(s,\theta\right)=-{\sqrt{s}\over \Gamma(n)}\rho(\theta){\cos\theta}{\left(2n\,\rho(\theta){-}n{+}1\right)_{n{-}1}} \, \pFq{2}{1}{ {{-}n\,\rho(\theta)}, 1{-}n}{ 1{-}2n\,\rho(\theta)}{z^{*}(\theta)}
\ee
with
\be\label{scat14} \rho(\theta) = \frac{1}{1+\sin\theta}\,,\qquad z^{*}(\theta) = \frac{1}{\cos^2(\theta/2)} \ee
The $s$ dependence in the dressing factor reduces to a simple prefactor of $s^{J/2}$.

\subsection{The HES amplitude in the Regge regime}
In the kinematical regime where $s\gg |t|$ one can study the small angle dependence of the amplitude, in particular the amplitude behavior in the Regge regime is provided by \eqref{ampint} specialized around the leading contribution at $z=1$. Alternatively it can be obtained from the representation \eqref{ampgen} in the limit $s\gg t$. From the latter one can recover a trivial action of the derivative operators which corresponds to the replacement of $d/d\beta\rightarrow1$. In any case the amplitude can be written as 
\be\label{ampgenReg}
{\cal A}_{gen}^{HES}(s,t)\Big|_{Regge}={\cal A}^{Regge}_{Ven}(s,t)\,e^{\sum_{n}{\cal J}_{n}{\cal O}_{n}\left(1\right)+\sum_{n,m}{\cal J}_{n}{\cal J}_{m}{\cal M}_{n,m}\left(1\right)}
\ee 
where there is the standard behavior of the Veneziano amplitude
\be\label{scat15}
{\cal A}^{Regge}_{Ven}(s,t) \sim \Gamma\Big({-}{t\over2}{-}1\Big) \,s^{{t\over2}+1}
\ee
and a non-trivial dressing factor made of the following contributions (see appendix B for details):
\begin{align}\label{scat16}
{\cal O}_{n}\left(1\right)=&\,\,(-)^{n}\zeta_{n}{\cdot}p_{1}  {(1{+}nq{\cdot}p_{1})_{n{-1}}\over \Gamma(n)}   \\
{\cal M}_{n,m}(1)=&\,\,{\zeta_{n}{\cdot}\zeta_{m}}{n m\over n{+}m}q{\cdot}p_{1}(1{+}q{\cdot}p_{1}) \,{ {\cal O}_{n}\left(1\right){\cal O}_{m}\left(1\right)\over  \zeta_{n}{\cdot}p_{1}\zeta_{m}{\cdot}p_{1}}    
\end{align}

Since $s\gg 2N$ and $s\gg |t|$ such that $\sin\theta\ll1$, one can take simplified forms for the kinematic factors in appendix \ref{app:4ptkin}, and see the explicit dependence on the angle. In particular one can take
\be\label{scat17}
q{\cdot}p_{1}=-{1\over 1+\sin\theta}\approx -1+ \sin\theta
\ee
to write
\begin{align}\label{scat18}
{\cal O}_{n}\left(1\right)\approx&-\zeta_{n}{\cdot}p_{1}{\Gamma(n\sin\theta)\Gamma(n{-}n\sin\theta)\over \Gamma(n)}\sin({n\pi\sin\theta}) \\
{\cal M}_{n,m}(1)\approx&\,\,{\zeta_{n}{\cdot}\zeta_{m}}{n m\over n{+}m}\sin\theta\,\,{ {\cal O}_{n}\left(1\right){\cal O}_{m}\left(1\right)\over  \zeta_{n}{\cdot}p_{1}\zeta_{m}{\cdot}p_{1}}
\end{align}
In the Regge regime, products of ${\cal M}_{n,m}(1)$ are suppressed since they contain higher powers of $\sin\theta$, allowing us to consider only the leading contributions provided by ${\cal O}_{n}\left(1\right)$. The ${\cal M}_{n,m}$ can also be eliminated by a choice of circular polarizations.

Finally, we can write the Regge limit of the amplitude
\be
{\cal A}^{HES}_{Regge}={\cal A}^{Regge}_{Ven}(s,t)\prod_{n=1}^{N}\big({\cal O}_n(1)\big)^{g_{n}} \label{eq:aHESRegge_O}
\ee
in an explicit form as
\be
{\cal A}^{HES}_{Regge}={\cal A}^{Regge}_{Ven}(s,t)\Big(-\sqrt{s}(1-\frac12\sin\theta)\Big)^{J}\prod_{n=1}^{N}\left({\Gamma(n\sin\theta)\Gamma(n{-}n\sin\theta)\over \Gamma(n)}\sin({n\pi\sin\theta})\right)^{g_{n}} \label{eq:aHESRegge}
\ee 
where it was used that $\zeta{\cdot}p_{1}\approx\sqrt{s}(1-\frac12\sin\theta)$ in this limit. The dressing factor of this amplitude bears strong resemblance to the three-point function of eq. \eqref{eq:A_HTT}.

\subsection{Spacing ratios for the four-point scattering amplitude}
We will search for chaotic behavior in the scattering angle, by analyzing the ratios of spacings of consecutive peaks for the amplitude derived above. The procedure will be the same as the one described in section \ref{sec:fitmodel}.

We have seen that the amplitude can be factorized as the Veneziano amplitude times a dressing factor that depends on the HES state which we write as
\be\label{scat21} {\cal A}_{HES}(s,t) = {\cal A}_{Ven}(s,t) {\cal D}_{HES}(s,\theta) \ee
The dressing factor ${\cal D}_{HES}$ is a complicated product of polynomials whose peaks are spaced erratically. We show that for solutions of
\be F_{\cal D}(\theta) \equiv \frac{d\log {\cal D}_{HES}}{d\theta} = 0 \label{eq:FD}\ee
the ratios of consecutive spacings will follow again the $\beta$-ensemble distribution, in both the fixed angle and Regge limits.

The interplay between the HES dressing factor and the Veneziano amplitude it multiplies can create a transition from chaotic to regular spacings, as will be discussed at the end of this section.

\subsubsection{Chaotic behavior in the fixed angle regime} \label{sec:chaosFA}
The dressing factor in the high energy limit was
\be\label{scat22} {\cal D}_{HES}^{f.a.}(s,\theta) = \prod_{n=1}^N\big({\cal T}^{(2)}_{n}(s,\theta) + {\cal T}^{(3)}_{n}(s,\theta) \big)^{g_n} \ee
where we can write the polynomial form of ${\cal T}^{(2)}_{n}$ and ${\cal T}^{(3)}_{n}$ in this limit, which explicitly reads
\begin{align}\label{scat23}
{\cal T}^{(2)}_{n}(s,\theta)=& \frac{\sqrt{s}}{\Gamma(n)}\left(\rho(\theta)\cos\theta{-}1\right)\sum_{k=0}^{n-1}\binom{n{-}1}{k}(2n\rho(\theta){-}n)_{n{-}k{-}1} (1 {-} n\rho(\theta))_{k} \left(\frac{1}{\cos^2(\theta/2)}\right)^{k+1} \nonumber \\
{\cal T}^{(3)}_{n}(s,\theta)=& \frac{\sqrt{s}}{\Gamma(n)} \rho(\theta)\cos\theta\sum_{k=0}^{n-1}\binom{n{-}1}{k}(2n \rho(\theta){-}n {+} 1)_{n-k-1} (-n \rho(\theta))_{k}  \left(\frac{1}{\cos^2(\theta/2)}\right)^{k}
\end{align}
with $\rho(\theta)=1/(1+\sin\theta)$. Since the $s$ dependence is only a prefactor in this limit it cannot affect the distribution of spacings.

\begin{figure}[ht!]
    \centering
    \includegraphics[width=0.48\textwidth]{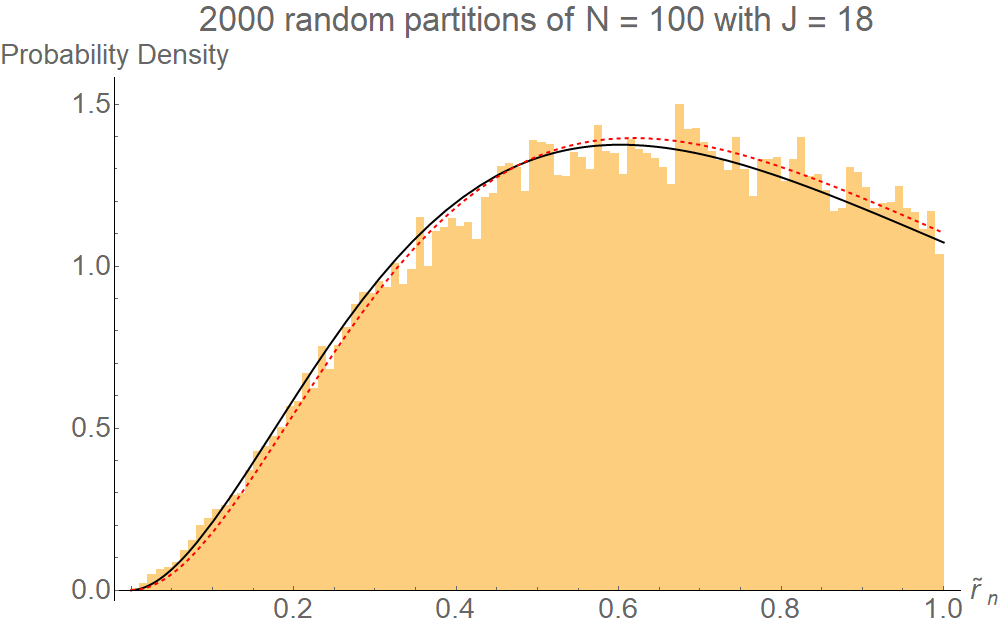}
    \caption{Distribution for $\tilde r$ for the dressing factor in the fixed angle limit. The best fit is with $\beta=1.86$ (black line), which is very close to the GUE (dashed red line). The average value is $\langle \tilde r_n \rangle = 0.600$.}
    \label{fig:r-FA-dressing}
\end{figure}

Taking a sample of 2000 random partitions of $N=100$, $J=18$, we collect all the solutions of $d\log{\cal D}/d\theta = 0$ in the range $\theta\in(0,\frac\pi2)$, and find the distribution plotted in figure \ref{fig:r-FA-dressing}. There are 29 data points per state in the sample.

We notice that there is an asymmetry in the distribution about $r\to1/r$ with values $r>1$ being favored, which causes it to deviate from the prediction. The average $\langle r_n\rangle $ is 1.497, while $\langle 1/r_n\rangle = 1.318$, where they should be equal. However, if one looks only at the normalized variable $\tilde r = \min(r,1/r)$, which in a way symmetrizes the distribution, the agreement with the $\beta$-ensemble is improved considerably. It is not clear what is the reason that $r>1$ is preferred, but the asymmetry could be a consequence of our looking solely at the part ${\cal A}(s,t)$ of the full amplitude, or of not looking at the full range of allowed angles.

\subsubsection{Chaotic behavior in the Regge regime} \label{sec:chaosRegge}
In the Regge regime, the dressing factor that depends on the HES state is
\be\label{scat24} {\cal D}_{HES}^{Regge}(s,\theta) = \Big({-}\sqrt{s}(1+\frac{\cos\theta}{1+\sin\theta})\Big)^{J}\prod_{n}\left(\frac{(1-n+\frac{n}{1+\sin\theta})_{n-1}}{\Gamma(n)}\right)^{g_n}
\ee
where in the present analysis we take eq. \eqref{eq:aHESRegge_O} before the small angle approximation. This is done mainly for practical reasons, since we have to go beyond small angles ($\sin\theta\ll 1$) to collect enough data points for the statistical analysis.\footnote{An alternative would be to use \eqref{eq:aHESRegge} and ignore the fact that we assumed $\sin\theta\ll1$ to get it. That dressing factor will also have a similar distribution of $r_n$.}

\begin{figure}[ht!]
    \centering
    \includegraphics[width=0.48\textwidth]{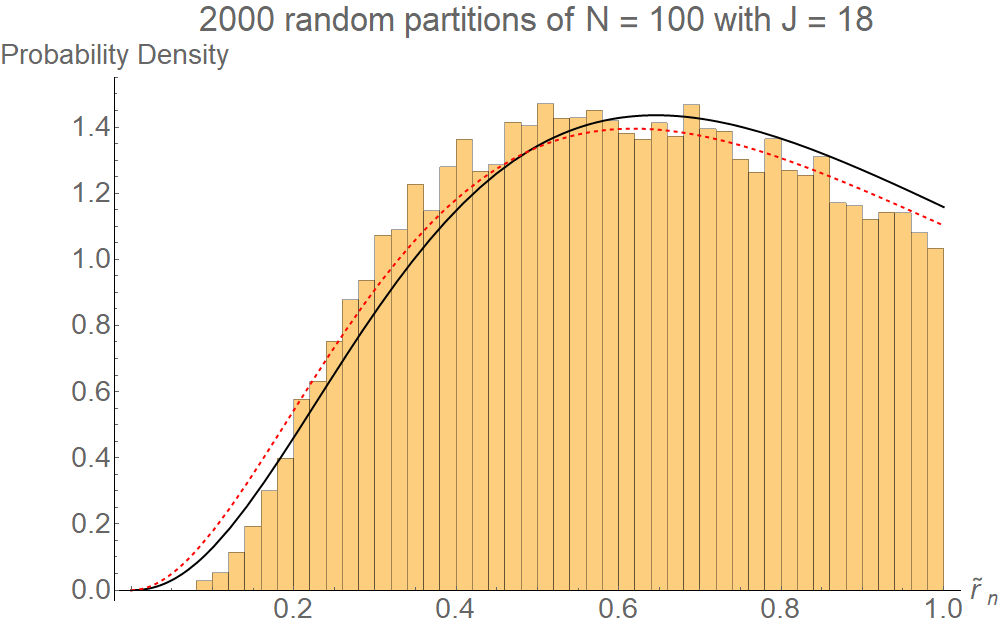}
    \caption{Distribution for $\tilde r$ for the dressing factor in the Regge limit. We take zeros in the range $\theta\in(0,\pi/4)$ and find 17 points per state. The average of $\langle \tilde r_n \rangle$ is 0.605 and the best fit (black) has $\beta = 2.27$. The distribution is also close to GUE (red, dashed).}
    \label{fig:r-Regge-dressing}
\end{figure}

It has almost the same form as the three point amplitude of eq. \eqref{eq:A_HTT} after the replacement of $\cos^2\frac\alpha2 \to 1/(1+\sin\theta)$. Each term in the product is a polynomial in $1/(1+\sin\theta)$.
The dependence of the dressing factor on $s$ is trivial, and the logarithmic derivative of the dressing factor is a function only of $\theta$ of a simple enough form
\be F_{\cal D}(\theta) = \frac{d\log {\cal D}_{HES}^{Regge}}{d\theta} =  -\frac{J}{1+\cos\theta+\sin\theta}-\frac{\cos\theta}{1+\sin\theta} \sum_{n=1}^N g_n\sum_{k=1}^{n-1}\frac{n}{n-k(1+\sin\theta)} \label{eq:dlogDR} \ee
We can find the zeros of this function of $\theta$ and plot the distribution of $\tilde r_n$ for their spacings, when considering only zeros in the range $\theta\in(0,\frac\pi4)$. The results are similar to what we obtained in the fixed angle regime in the previous subsection, with a slightly skewed distribution that agrees well with the $\beta$-ensemble distribution once we symmetrize it. See figure \ref{fig:r-Regge-dressing}.

\subsubsection{Transition from chaotic to regular behavior}
The interplay between the Veneziano amplitude and its dressing factor can create a transition from chaotic to regular spacings as one moves from small to large angles.

The erratic dressing factor is multiplied by the Veneziano amplitude, which is a highly oscillatory function of the angle at high energies - but one that is regular in the sense that it has regularly spaced zeros as a function of $t$, which leads to almost regularly spaced peaks of the amplitude between these zeros. As a function of $\theta$ these zeros become denser at large angles, but remain quite regular. In the interplay between the Veneziano and the HES dressing factors, the spacings between peaks depend on which of the two is oscillating faster. For large angles, the Veneziano factor will usually dominate and cause the peaks to be regularly spaced, while at smaller angles one can see the chaotic spacings coming from the HES factor.

Whether we can see these two regimes clearly will depend on the scattering energy, and on the level of the HES state. It is somewhat difficult to find the range of parameters ($s$ and $N$) where this can be computed and seen clearly. One example in which we see it in the Regge limit amplitude of \eqref{eq:aHESRegge}.\footnote{This is not an optimal example, since we should not trust \eqref{eq:aHESRegge} away from small angles. We choose it for illustration purposes, and because it is the simplest amplitude that we can write that realizes this transition in an explicit and easy to calculate way.}

Recall that the full form of the Veneziano amplitude in the Regge limit is
\be\label{scat25} {\cal A}_{Ven}^{Regge}(s,t) = \frac{\sin[\pi \alp (s+t)]}{\sin(\pi \alp s)} \Gamma(-\alp t-1) (\alp s)^{\alp t +1} \ee
Importantly, here we retain the oscillating factor which is usually averaged over in discussing the Regge behavior. Taking the log derivative we get the term
\be\label{scat26} \frac{d\log {\cal A}_{Ven}^{Regge}}{d\theta} = \frac{d t}{d\theta} \alp \left(\log(\alp s) + \pi \cot[\pi\alp(s+t)] - \psi(-\alp t-1)\right)\ee
which is a complicated function of the angle with many zeros, which are almost regularly spaced. Note the implicit dependence on the level $N$ which enters through the kinematic relation of $t(\theta)$.

\begin{figure}[ht!]
    \centering
    \includegraphics[width=0.48\textwidth]{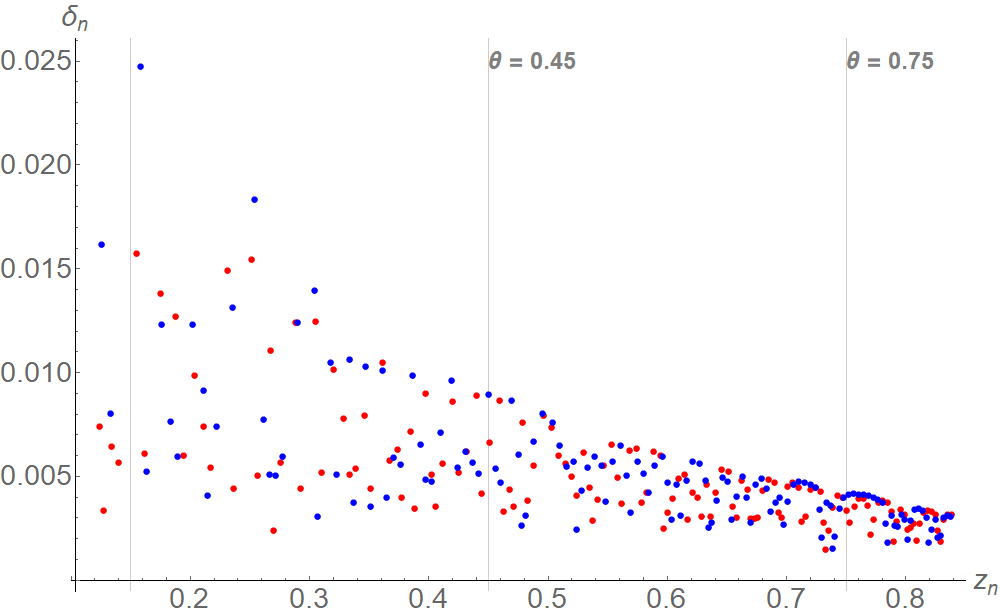}\\
    \includegraphics[width=0.48\textwidth]{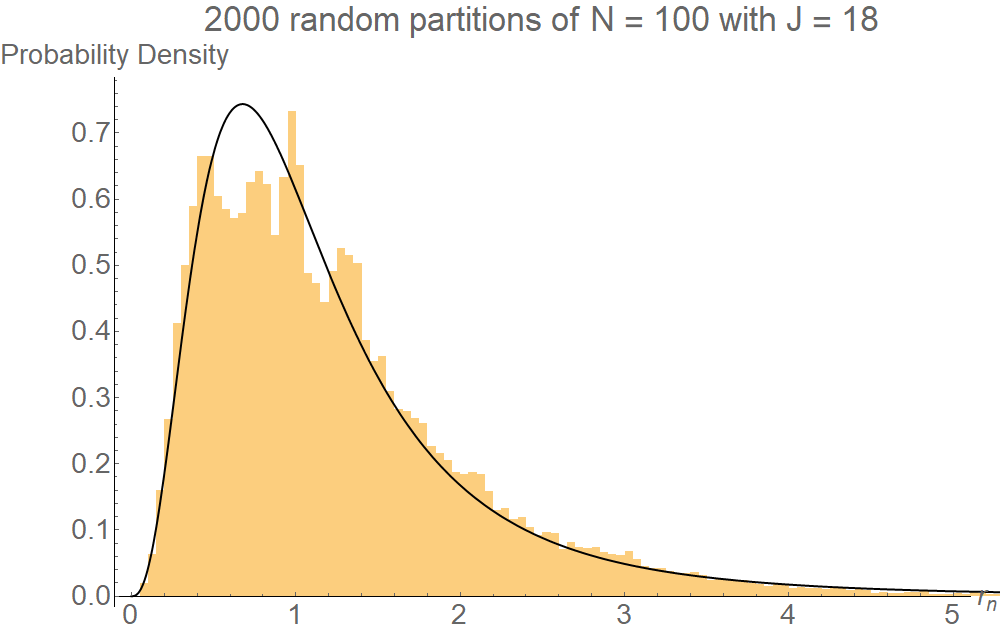}
    \includegraphics[width=0.48\textwidth]{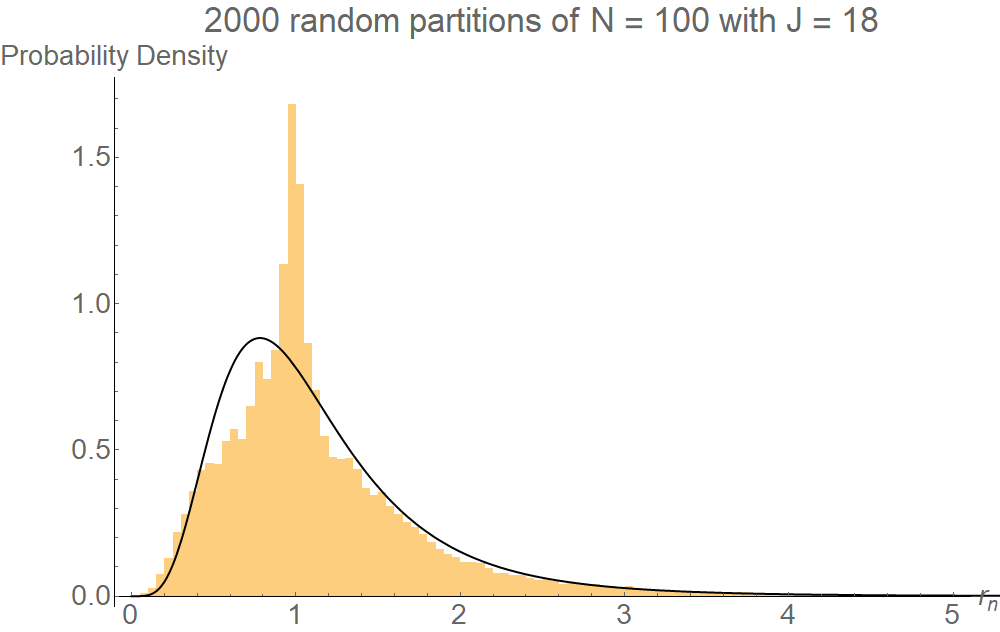}
    \caption{Top: The spacings $\delta_n$ as a function of $z_n$  for two random states of $N=100$. The spacings exhibit a transition from random to regular behavior. Bottom: The distribution of spacing ratios ($r_n$) for zeros in the ranges $\theta\in(0.15,0.45)$ (left) and $\theta\in(0.15,0.75)$ (right). In the latter, a narrow peak around $r=1$ appears on top of the chaotic distribution.}
    \label{fig:transition}
\end{figure}

Now, in studying the distribution of zeros of the log-derivative of the full amplitude,
\be F(\theta) = \frac{d\log {\cal A}_{Ven}^{Regge}}{d\theta} + \frac{d\log {\cal D}_{HES}^{Regge}}{d\theta} = 0 \label{eq:F4Regge} \ee
we can observe a transition between the chaotic and regular spacings, as illustrated in figure \ref{fig:transition}.

\section{Summary and outlook} \label{sec:summary}
We have expanded on the analysis of string scattering amplitudes involving an HES state, randomly chosen among the huge number of states with the given mass ($\alp  M^2 {=}  N{-}1 {=}\sum_n n g_n-1$) and helicity $J$ ($N{=}S_{Max} \ge S \ge J {=}\sum_n g_n$).  In \cite{Bianchi:2022mhs} we fitted the spacing ratios to a log-normal distribution. Although the log-normal distribution proved to be a good approximation of the available results, in the present work we have improved and argued that the $\beta$-ensemble distribution represents a better fit for the larger and finer set of data analyzed here. Moreover we have also argued that the $\beta$-ensemble distribution is better motivated by physical considerations.

We observed the chaotic behavior in the angular dependence of scattering amplitudes involving one highly excited string state. We can ascribe the chaos to the random superposition of the relevant spherical harmonics in the partial wave expansion of the amplitude. Given the unitarity of the process (the overall normalization is irrelevant in the analysis) we would expect $\beta$-ensemble distribution with $\beta=2$ as for GUE. Instead we found better fits with values of $\beta$ not far from 2 but significantly different from it such that the distribution cannot be precisely that of GUE.

The scattering amplitude of HES states and three tachyons which we derived and analyzed in various regimes shows chaotic behavior associated to a dressing factor multiplying the 'standard' Veneziano amplitude. In this case we did not examine in as much detail the dependences of the distributions on $N$ and $J$, but our observation is that the general behavior is the same as in the three-point function.

There are several possible directions for followup work: 
\begin{itemize}
\item One should try and clarify the origin of the mild but non-negligible dependence of the fitted $\beta$ parameter and the average ratio $\langle r \rangle$ on the chosen data, most prominently on $N$ and $J$.

\item One could conceive extending the analysis to processes with two or more HES states. The first obvious choice that comes to mind is generalized Compton scattering of a low-mass probe off HES states, possibly including inelastic channels involving excitations of the HES itself (to a nearby HES state). A related amplitude was computed already in \cite{Hashimoto:2022bll}.

\item One could consider similar processes in hadronic models inspired by holographic QCD, along the lines of \cite{Polchinski:2001tt, Bianchi:2021sug} and prove or disprove the emergence of chaotic behavior in this context.

\item Finding other tractable examples for chaos in quantum scattering amplitudes is another important task. In \cite{Bianchi:2022mhs} we have discussed the leaky torus as another example where an analytic formula is known \cite{Gutzwiller:1983}. Another example of chaotic quantum scattering is in \cite{Fukushima:2022lsd}.

\item One could give further quantitative support to the string/black hole correspondence by relating the ``measurable'' chaotic behavior found here with the expected chaotic behavior of scattering, absorption, and thermalization in black hole processes \cite{Bianchi:2020des, Bianchi:2020yzr, Bacchini:2021fig, Berti:2009kk, Aminov:2020yma, Bianchi:2021mft, Bonelli:2021uvf, Bianchi:2021xpr, Bianchi:2022qph, Shenker:2013pqa, Polchinski:2015cea, Martinec:2020cml}. To this end, a given HES state, that may be considered a (a-)typical microstate of a putative black hole ensemble, would better be replaced with a coherent state with mass, charge and spin distributed around some large ``classical'' value. Using the results in \cite{Bianchi:2019ywd} could prove crucial in this endeavour.

\item In \cite{Srdinsek:2020bpq}, it was observed that while the spectra of some non-integrable QFT models displayed chaotic behavior, the associated distribution of eigenvectors did not match the RMT expectations, and was in fact unchanged from the integrable, non-chaotic models. One should identify what are the eigenvectors associated with the scattering amplitudes and verify if they are distributed as predicted by RMT. This has consequences for thermalization, as stressed in \cite{Srdinsek:2020bpq}.
\end{itemize}

We hope to report soon on some of the above issues.

\section*{Acknowledgments}
We would like to thank R.~Benzi, B.~Craps, K.~Hashimoto, G.~Parisi, G.~Salina, F.~Popov, D.~Gross, S.~Negro, V.~Rosenhaus, S.~Yankielowicz, P.~Di~Vecchia, B.~Sundborg, E.~Kiritsis, V.~Niarchos and T.~Yoda for useful discussions. The work of M.~B. and M.~F. is partially supported by the MIUR PRIN Grant 2020KR4KN2 ``String Theory as a bridge between Gauge Theories and Quantum Gravity''. The work of J.~S. was supported by a
center of excellence of the Israel Science Foundation (grant number 2289/18). The work of D.~W. was supported by the Quantum Gravity Unit of the Okinawa Institute of Science and Technology Graduate University (OIST) when this project was begun, and by the Young Scientist Training (YST) Program of the Asia Pacific Center for Theoretical Physics (APCTP) when it was completed.

\appendix

\section{Four-point amplitude kinematics}\label{app:4ptkin}
Let us introduce the center of mass system (CMS) kinematics of scattering three tachyons of momenta $p_1$, $p_2$ and $p_3$ with an HES state of momentum $p$.

We choose the momenta of the scattered states as 
\be\label{fff1}
p_{1}=(E_{1},p_{in},0,\vec{0}),\,\,\,p_{2}=(E_{2},-p_{in},0,\vec{0})
\ee
\be\label{fff2}
p_{3}=-(E_{3},p_{out} \cos\theta,p_{out} \sin\theta,\vec{0}),\,\,\,p=-(E_{4},-p_{out} \cos\theta, -p_{out} \sin\theta,\vec{0})
\ee
such that
\be\label{fff3}
p_{1}+p_{2}=\sqrt{s}\,, \quad  p_{1}+p_{2}+p_{3}+p=0
\ee 

The DDF reference light-like momentum $q$ and polarization $\lambda$ are constrained by $q{\cdot}p=1$\footnote{Here and throughout we work in units where $\alp = \frac12$.} and $\lambda{\cdot}q=0$, and they are represented by
\be\label{fff4}
q={(1,0,1,\vec{0})\over E_{4}+\sin\theta\,p_{out}}\,,\,\,\, \lambda={(0,1,0,\vec{\Lambda})\over \sqrt{1+\vec{\Lambda}{\cdot}\vec{\Lambda}^*}}
\ee
Since, for simplicity, we chose to work with ``null'' $\lambda$, $\lambda{\cdot}\lambda=0$, we may choose a complex $\vec{\Lambda}$ such that $\vec{\Lambda}{\cdot}\vec{\Lambda}=-1$ and yet $\vec{\Lambda}{\cdot}\vec{\Lambda}^*=+1$.

Following the CMS kinematics one has
\be\label{fff4b}
E_{1}={s+M_{1}^{2}-M_{2}^{2}\over 2 \sqrt{s}}={\sqrt{s}\over2}\,,\,\,\, E_{2}={s+M_{2}^{2}-M_{1}^{2}\over 2 \sqrt{s}}={\sqrt{s}\over2}
\ee
\be\label{fff4c}
E_{3}={s+M_{3}^{2}-M_{4}^{2}\over 2 \sqrt{s}}={s-2N\over 2 \sqrt{s}}\,,\,\,\,E_{4}={s+M_{4}^{2}-M_{3}^{2}\over 2 \sqrt{s}}={s+2N\over 2 \sqrt{s}}
\ee
and
\be\label{fff6}
p_{in}^{2}=2+{s \over 4 }\,; \quad p_{out}^{2}=2+ {s\over 4}\left(1-{2N\over s}\right)^{2}
\ee

The relevant scalar products involving momenta are given by
\be\label{fff7}
q{\cdot}p_{1}=-{E_{1}\over  \sin\theta\,p_{out}+E_{4}}={-1\over 1+{2N\over s}+ 2  \sin\theta \sqrt{{2\over s}+ {1\over 4}\left(1-{2N\over s}\right)^{2}} }=q{\cdot}p_{2}
\ee
\be\label{fff8}
q{\cdot}p_{3}={E_{3}-p_{out}\sin\theta\over E_{4}+p_{out}\sin\theta }={1-{2N\over s}-  2\sin\theta \sqrt{{2\over s}+ {1\over 4}\left(1-{2N\over s}\right)^{2}}\over 1+{2N\over s}+  2\sin\theta \sqrt{{2\over s}+ {1\over 4}\left(1-{2N\over s}\right)^{2}} }
\ee
Using conservation of momentum and $q\cdot p = 1$ one can easily derive the relation
\be\label{fff9} q{\cdot}p_3 = - 1 - q{\cdot}(p_1+p_2) = - 1 - 2 q{\cdot}p_1 \ee

The relevant scalar products involving polarizations are given by
\be\label{fff10}
\lambda{\cdot}p=p_{out}\cos\theta=\sqrt{s}\cos\theta \sqrt{{2\over s}+ {1\over 4}\left(1-{2N\over s}\right)^{2}}=-\lambda{\cdot}p_{3}
\ee
\be\label{fff11}
\lambda{\cdot}p_{1}=p_{in}=\sqrt{s}\,\sqrt{{2\over s}+{1 \over 4 }}=- \lambda{\cdot}p_{2}
\ee
where for convenience it was constrained the free parameter to be $\vec{\Lambda}{\cdot}\vec{\Lambda}{=}-1$. Given the general form of the covariant polarization $\zeta^{\mu}=\lambda^{\mu}-\lambda{\cdot}p\,q^{\mu}$, it follows that 
\be\label{fff12}
\zeta{\cdot}p_{1}=\sqrt{s}\,\sqrt{{2\over s}+{1 \over 4 }} + {\sqrt{s}\cos\theta\sqrt{{2\over s}+ {1\over 4}\left(1-{2N\over s}\right)^{2}} \over  1+{2N\over s}+ 2  \sin\theta \sqrt{{2\over s}+ {1\over 4}\left(1-{2N\over s}\right)^{2}}  }
\ee
\be\label{fff13}
\zeta{\cdot}p_{2}=-\sqrt{s}\,\sqrt{{2\over s}+{1 \over 4 }}+ {\sqrt{s}\cos\theta\sqrt{{2\over s}+ {1\over 4}\left(1-{2N\over s}\right)^{2}} \over  1+{2N\over s}+ 2  \sin\theta \sqrt{{2\over s}+ {1\over 4}\left(1-{2N\over s}\right)^{2}}  }
\ee
\be\label{fff14}
\zeta{\cdot}p_{3}=- {2\sqrt{s}\cos\theta\sqrt{{2\over s}+ {1\over 4}\left(1-{2N\over s}\right)^{2}} \over  1+{2N\over s}+ 2  \sin\theta \sqrt{{2\over s}+ {1\over 4}\left(1-{2N\over s}\right)^{2}}  }
\ee
the combination of these terms reflects momentum conservation $\zeta{\cdot}(p_{1}+p_{2}+p_{3})=0$ and transversality $\zeta{\cdot}p{=}0$, as expected.

\section{Derivation of the HES-three tachyon amplitude} \label{app:4ptderivation}
In this section we will review the main steps for the computation of the scattering amplitude involving three tachyons and one generic HES state in open bosonic string. Let us start by considering the tachyonic vertex operators inserted in the ordered positions $z_{j}$ of the disk, with $j=1,2,3$
\be\label{fff14b}
V_{T}(p_{j},z_{j})=e^{ip_{j}{\cdot}X}(z_{j})
\ee
and the coherent vertex operator inserted in position $z_{4}$
\be\label{cohez4}
V_{{\cal C}}(p,z_{4})=\exp{\left(\sum_{n,m}{\zeta_{n}{\cdot}\zeta_{m}\over 2}{\cal S}_{n,m}e^{-i(n{+}m)q{\cdot}X}{+}\sum_{n}\zeta_{n}{\cdot}{\cal P}_{n}e^{-inq{\cdot}X}{+}i\widetilde{p}{\cdot}X\right)}(z_{4})
\ee
The generating scattering amplitude is given by
\be\label{fff15}
{\cal A}_{gen}^{HES}(s,t)=\int_{z_{4}}^{z_{2}} \prod_{\ell=1}^{4}dz_{\ell}\,\Big\langle V_{T}(p_{1},z_{1})\,V_{T}(p_{2},z_{2})\,V_{T}(p_{3},z_{3})\,V_{{\cal C}}(p,z_{4}) \Big\rangle
\ee
From the correlator one can factorize the Koba-Nielsen contribution due to the contractions 
\be\label{fff16}
\langle p_{j}{\cdot}X(z_{j})\,p_{\ell}{\cdot}X(z_{\ell})\rangle=-\,p_{j}{\cdot}p_{\ell}\log(z_{j\ell}) \,, \quad z_{j\ell}=z_{j}{-}z_{\ell}
\ee
finding 
\be\label{fff16b}
KN(\{z_{j}\})=z_{12}^{p_{1}{\cdot}p_{2}}z_{13}^{p_{1}{\cdot}p_{3}}z_{14}^{p_{1}{\cdot}p}z_{23}^{p_{2}{\cdot}p_{3}}z_{24}^{p_{2}{\cdot}p}z_{34}^{p_{3}{\cdot}p}
\ee
Using the general kinematics
\be\label{fff17}
(p_{3}{+}p)^{2}=-s=(p_{1}{+}p_{2})^{2}\,,\,\,\,(p_{2}{+}p_{3})^{2}=-t=(p_{1}{+}p)^{2}\,,\,\,\, (p_{1}{+}p_{3})^{2}=-u=(p_{2}{+}p)^{2}\,,
\ee
where $s+t+u=3M_{T}^{2}+M_{N}^{2}$ with $M_{N}^{2}=2(N{-}1)$ the mass square of a generic state of the level $N$, the Koba-Nielsen contribution can be written as
\be\label{fff18}
KN(\{z_{j}\})=\left(z_{12}z_{34}\over z_{13}z_{24} \right)^{-{s\over 2}{-}2}\left(z_{14}z_{23}\over z_{13}z_{24} \right)^{-{t\over 2}{-}2} (z_{13}z_{24})^{{-}2}\, \left({z_{34}z_{14}\over z_{13}} \right)^{N}
\ee
The last factor is related to the exponential nature of the insertion (\ref{cohez4}), and it can be reabsorbed as a dressing factor of the linear and bilinear contributions 
\be\label{fff19}
e^{-inq{\cdot}X}\Rightarrow  \left({z_{34}z_{14}\over z_{13}} \right)^{n} \,, \quad e^{-i(m+n)q{\cdot}X}\Rightarrow  \left({z_{34}z_{14}\over z_{13}} \right)^{m+n}
\ee 

The remaining contributions are related to the contractions
\be\label{fff20}
\langle p_{j}{\cdot}X(z_{j})\,\zeta_{n}{\cdot}\partial X(z_{4}) \rangle={\zeta_{n}{\cdot}p_{j}\over z_{j4}}
\ee
that combined with the operator structure $\zeta_{n}{\cdot}{\cal P}_{n}$ and ${\cal S}_{n,m}$ yield 
\be\label{fff21}
\zeta_{n}{\cdot}{\cal P}_{n}(z_{4})\Rightarrow\, \sum_{k=1}^{n}\left(\prod_{j\ne4}{\zeta_{n}{\cdot}p_{j}\over z_{j4}}\right){\cal Z}_{n-k}\left(n\sum_{j\ne4}{q{\cdot}p_{j}\over z_{j4}} \right)
\ee
and
\be\label{fff22}
{\cal S}_{n,m}(z_{4})=\sum_{r=1}^{m}r\,{\cal Z}_{n{+}r}\left(n\sum_{j\ne4}{q{\cdot}p_{j}\over z_{j4}} \right){\cal Z}_{m{-}r}\left(m\sum_{j\ne4}{q{\cdot}p_{j}\over z_{j4}} \right)
\ee 
The combination of all the contractions yields
\be\label{fff23}
\begin{split}
{\cal A}_{gen}^{HES}(s,t)=&\int_{z_{4}}^{z_{2}} \prod_{\ell=1}^{4}dz_{\ell}\,\left(z_{12}z_{34}\over z_{13}z_{24} \right)^{-{s\over 2}{-}2}\left(z_{14}z_{23}\over z_{13}z_{24} \right)^{-{t\over 2}{-}2} (z_{13}z_{24})^{{-}2}\\
&\exp{\left\{\sum_{n} \left({z_{34}z_{14}\over z_{13}} \right)^{n} \sum_{k=1}^{n}\left(\prod_{j\ne4}{\zeta_{n}{\cdot}p_{j}\over z_{j4}}\right){\cal Z}_{n-k}\left(n\sum_{j\ne4}{q{\cdot}p_{j}\over z_{j4}}\right) \right\}}\\
&\exp{\left\{ \sum_{n,m}{\zeta_{n}{\cdot}\zeta_{m}\over 2} \left({z_{34}z_{14}\over z_{13}} \right)^{n{+}m} \sum_{r=1}^{m}r\,{\cal Z}_{n{+}r}\left(n\sum_{j\ne4}{q{\cdot}p_{j}\over z_{j4}} \right){\cal Z}_{m{-}r}\left(m\sum_{j\ne4}{q{\cdot}p_{j}\over z_{j4}} \right)\right\}}
\end{split}
\ee
Eliminating the redundancy of the $SL(2,\mathit{R})$ invariance fixing $z_{1},z_{2}$ and $z_{4}$ produces the standard transformation of the integration measure
\be\label{fff24}
\prod_{j=1}^{4}dz_{j}=z_{12}z_{14}z_{24}\, dz_{3}
\ee
which combined with $(z_{13}z_{24})^{-2}$ gives a finite result in the limit $z_{1}{=}\infty,\,z_{2}{=}1,\,z_{3}=z,\,z_{4}{=}0$   
\be\label{fff25}
{z_{12}z_{14}\over z_{13}^{2}z_{24} }\Big|_{z_{1}=\infty}=1
\ee 
while for the other terms one has
\be\label{fff26}
\left(z_{12}z_{34}\over z_{13}z_{24} \right)^{-{s\over 2}{-}2}=z^{-{s\over 2}{-}2}\,, \quad  \left(z_{14}z_{23}\over z_{13}z_{24} \right)^{-{t\over 2}{-}2}=(1{-}z)^{-{t\over 2}{-}2}
\ee
Finally one can study how the remaining contributions transform under the $SL(2,\mathit{R})$ invariance, finding the final expression of the scattering amplitude given in \eqref{ampint}.

\section{Comparison of the \texorpdfstring{$\beta$}{beta}-ensemble with a log-normal distribution} \label{app:lognormal}
In our previous paper \cite{Bianchi:2022mhs} we fitted the spacing ratios $r_n$ to a log-normal distribution
\be f_{\text{LN}}(r) = \frac{1}{\sqrt{2\pi \sigma^2} r} \exp\left(-\frac{[\log(r)-\mu]^2}{2\sigma^2}\right) \label{eq:pdfLN} \ee
This distribution proved good as a first approximation of the result, but in the present work we have seen that the $\beta$-ensemble distribution \eqref{eq:beta_r} is both a better fit for the data, and is better motivated by random matrix theory.

The distributions are quite close. For a given $\beta$ we can find a log-normal distribution with $\mu = 0$ (because of the $r\to1/r$ symmetry) and an appropriate value of $\sigma$ to approximate it very well. One can measure the distance between the distributions using the relative entropy,
\be I(\beta;\sigma) = \int_{0}^\infty dr f_\beta(r) \log\left(\frac{f_\beta(r)}{f_{LN}(r)}\right) \ee
and, for a given $\beta$, pick the value of $\sigma$ that minimizes it. For instance, for $\beta = 1$, 2, and 4 the closest log-normal distributions are at $\sigma \approx 1.014$, $0.773$ and $0.565$ respectively. The value of the relative entropy at the minimum is in these three cases between $0.002$--$0.009$, with the better agreement occurring for larger $\beta$.

\begin{figure}[ht!]
    \centering
    \includegraphics[width=0.48\textwidth]{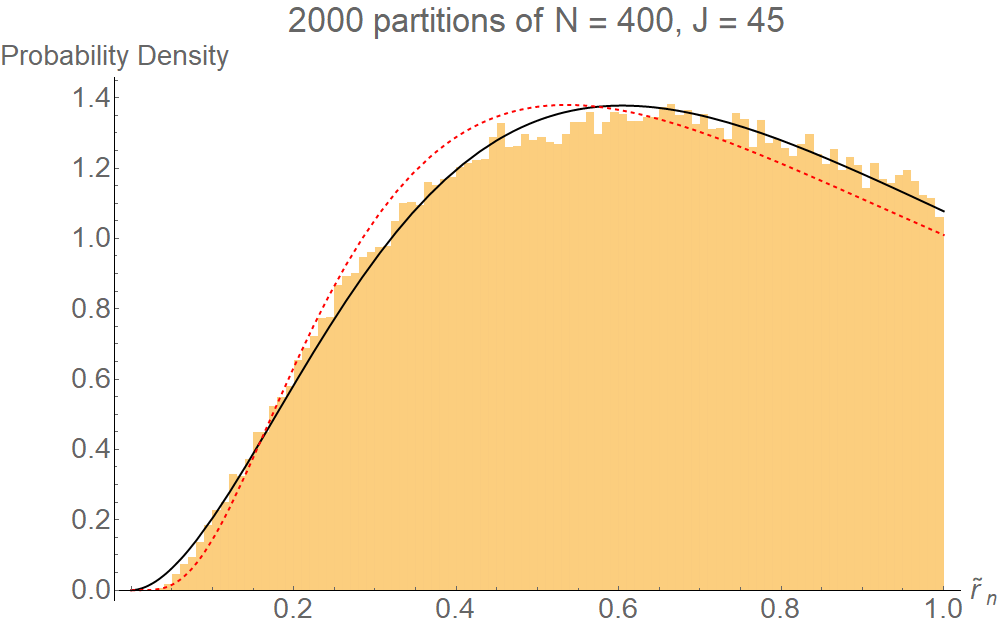}
    \caption{Distribution of spacing ratios in the decay amplitude of states with $N=400$, fitted to a log-normal distribution with $\sigma=0.79$ (dashed red line) and to the $\beta$-ensemble (black line) distribution with $\beta = 1.88$. The better fit is with $\beta$.}
    \label{fig:log-normal}
\end{figure}

For a direct comparison, in figure \ref{fig:log-normal} we plot one of the distribution we had in section \ref{sec:3pt}, for the decay amplitude of states with $N=400$, now together with a fitted log-normal distribution. We would also note that in \cite{Bianchi:2022mhs} we used an approximated form of the $HES\to TT$ amplitude \eqref{eq:A_HTT} which separately caused deviations in the result for the spacing ratios. The excess of points around $r=1$ noted there does not occur when we use the exact formula.

\section{Random partitions of a large integer} \label{app:randompartitions}
As discussed in section \ref{sec:partitions}, the number of partitions of an integer $N$ grows exponentially in $\sqrt{N}$. Since we cannot probe the full space of states, we need a reliable method of picking representative, generic states in a random way.

Picking a partition of a large integer $N$ at random, with each partition having an equal probability of being chosen, is a non-trivial task. We present here one algorithm that accomplishes this goal.

We represent a partition as a list $\{g_n\}$, $n = 1,2,\ldots N$, where $g_n$ is the number of times that $n$ occurs in the partition. The algorithm relies on a result \cite{Fristedt:1993} regarding the asymptotic distributions of $\{g_n\}$ for large $N$, namely that each $g_n$ has the geometric distribution
\be P(g_n = k) = (1-p_n)^k p_n \label{eq:distnm} \ee
with the parameter
\be p_n = 1 - \exp\left(-\frac{ n \pi}{\sqrt{6 N}}\right) \ee
One can generate a random partition of $N$ by drawing values of $\{g_n\}$, $n=1,2,\ldots,N$ from the above distribution, treating each $g_n$ as an independent variable, until one reaches a set that corresponds to a partition of $N$. That is, until one gets a set of $\{g_n\}$ that satisfies $\sum_n n g_n = N$. The result of \cite{Fristedt:1993} implies that the partitions of $N$ that will be reached by this algorithm will be uniformly distributed, at least at large $N$. Each partition of $N$ has an equal probability of being chosen.

The downside of the algorithm is that it needs to reject many sets of $\{g_n\}$ until it reaches one that satisfies the constraint, with the expected number of rejections being ${\cal O}(N^{3/4})$. By use of probabilistic algorithms one can improve the number of rejections to ${\cal O}(N^{1/4})$ or even ${\cal O}(1)$ \cite{Arratia:2016}.

The simpler, ${\cal O}(N^{1/4})$ algorithm is as follows:
\begin{enumerate}
    \item Draw $\{g_n\}$ for $n \geq 2$, with $g_n$ distributed according to \eqref{eq:distnm}.
    \item Set $k \equiv N - \sum_{n=2}^N n g_n$. If $k < 0$, restart from step 1.
    \item Draw a random variable $u \in (0,1)$ from the uniform continuous distribution. If $u < e^{-\frac{k \pi}{\sqrt{6N}}}$, reject the partition and return to step 1.
    \item Else, set $g_1 = k$ to finish.
\end{enumerate}
Step 3, where some partitions are rejected with a specifically chosen probability, assures that the probability to output any given partition is as before.

We can use a modification of the above algorithm to generate a partition of a given length $J$. We modify only step 1, where we start by choosing $\{g_n\}$ such that $g_J \geq 1$ and $g_{n>J} = 0$. Then, the result after step 4 will be a partition of $N$ where the maximum summand in the partition is $n_{\text{max}}=J$. Then, taking the conjugate partition, we get a partition of $N$ into exactly $J$ parts. 

In the preparation of this work, we have used several methods of picking random partitions. One is the brute force method: generate a list of all possible partitions of a given $N$ (and $J$ when that is constrained), then, select random elements from the list with equal probability. This is the simplest method at smaller $N$, but becomes impractical quickly as one increases $N$. For unconstrained partitions of $N$ we have used Mathematica's built-in  (as part of the Combinatorica package) function RandomPartition[$N$]. To produce partitions of large $N$ with fixed $J$, we have used the algorithm described in the previous paragraph.

\bibliographystyle{JHEP}
\bibliography{SACS}

\end{document}